\documentclass[aps,pra,floatfix,twocolumn,tigthenlines]{revtex4-1}

\usepackage{amsmath}
\usepackage{amssymb}
\usepackage{graphicx}
\usepackage{algorithm}
\usepackage{algorithmic}
\usepackage{units}

\usepackage{color} 
\usepackage[normalem]{ulem}

\newcommand{\ket}[1]{|#1 \rangle}

\begin{document}

\title{Photonic architecture for scalable quantum information processing  in NV-diamond}

\author{Kae Nemoto$^{1}$,  Michael Trupke$^{2}$, Simon J. Devitt$^{1}$,  Ashley M. Stephens$^{1}$, Kathrin Buczak$^{2}$, Tobias N\"obauer$^{2}$, Mark S. Everitt$^{1}$,  J\"org Schmiedmayer$^{2}$ \& William J. Munro $^{3,1}$}

\affiliation{$^1$ National Institute of Informatics, 2-1-2 Hitotsubashi, Chiyoda-ku, Tokyo 101-8430, Japan \\
$^2$ Vienna Center for Quantum Science and Technology, Atominstitut, TU Wien, 1020 Vienna, Austria \\
$^3$ NTT Basic Research Laboratories, NTT Corporation, 3-1 Morinosato-Wakamiya, Atsugi, Kanagawa 243-0198, Japan}

\begin{abstract} 
Physics and information are intimately connected, and the ultimate information processing devices will be those that harness the principles of quantum mechanics. Many physical systems have been identified as candidates for quantum information processing, but none of them are immune from errors. The challenge remains to find a path from the experiments of today to a reliable and scalable quantum computer. Here, we develop an architecture based on a simple module comprising an optical cavity containing a single negatively-charged nitrogen vacancy centre in diamond. Modules are connected by photons propagating in a fiber-optical network and collectively used to generate a topological cluster state, a robust substrate for quantum information processing. In principle, all processes in the architecture can be deterministic, but current limitations lead to processes that are probabilistic but heralded. We find that the architecture enables large-scale quantum information processing with existing technology.
\end{abstract}
\date{\today}

\maketitle

\section{Introduction} 
Quantum computers promise to surpass even the fastest classical computers, but the task of building a quantum computer presents a significant challenge. Even if they are precisely engineered, quantum systems will inevitably suffer from decoherence and other errors. If these errors are sufficiently rare and not too strongly correlated, then they can be suppressed with quantum error correction \cite{Devitt2009}. The role of quantum computer architecture is to integrate quantum error correction with feasible experimental technology, to find a path to a reliable and scalable quantum computer. In this context, of the many physical systems identified as candidates for quantum information processing \cite{Ladd2010}, the negatively-charged nitrogen vacancy (NV$^-$) centre in diamond \cite{Davies1976,Harley1984,Reddy1987} features a number of desirable properties \cite{Childress2005,Childress2006,Benjamin2006,Jiang2007,YJG12}. For example, the NV$^-$ centre possesses both a nuclear spin and an electron spin---the nuclear spin can serve as a memory to store quantum information for relatively long times \cite{Maurer2012}, and the electron spin can be coupled to a photon to serve as a flexible interface with other NV$^-$ centres \cite{Togan2010}. The experimental feasibility of this system has been well established in recent years. Experiments have demonstrated individual electron and nuclear spin initialisation, manipulation, and measurement \cite{Jelezko2004,Dutt2007,Neumann2008,Hanson2008,Jiang2009,Neumann2010,Neumann2010b,Buckley2010,Robledo2011,vanderSar2012,Dolde2013}, long-lived nuclear memories \cite{Maurer2012}, a coherent interface between an electron spin and an optical field \cite{Togan2010}, and optical cavities containing NV$^-$ centres \cite{Park2006,Englund2010,Faraori2011}. State-dependent reflectivity has been demonstrated with atoms \cite{Englund2007}, though not yet with NV$^-$ centres. At the same time, new techniques for quantum error correction have lessened experimental requirements \cite{Knill2005,Bacon2006,Raussendorf2007}.

Here, we develop a quantum computer architecture based on a simple module comprising an optical cavity containing a single NV$^-$  centre in diamond. Modules are connected by photons propagating in a fiber-optical network. The cavities mediate interactions between the photons and the electron spins, enabling entanglement distribution and readout. The electron spins are coupled to nuclear spins, which constitute long-lived quantum memories where quantum information is stored and processed. Aside from modules connected by optical fibers, other elements of the architecture are single-photon detection devices and classical control lines. These elements are laid out in a regular two-dimensional array, with sufficient connectivity between modules to enable topological cluster-state error correction \cite{RHG07, FG09, Barrett2010}. This arrangement is independent of the size of the network. At a circuit level, we find the maximum tolerable error per elementary quantum gate to be approximately 0.73\%. However, by analysing the architecture at the physical level, we also estimate how well each component of the module must operate for the system to meet this threshold and be truly scalable. The results of this analysis indicate that the architecture is consistent with present technology and might be achievable in the near future.

\section{Fundamental building blocks} 
Our approach can be adapted to a variety of promising physical systems, such as ions, neutral atoms, and quantum dots \cite{Togan2010,Stute2013,Greve2012,Schaibley2013,Gao2012}, and for this reason, we begin with a general description of the fundamental module. However, to show that the module can form the basis of a truly scalable architecture, we focus on a concrete implementation using NV$^-$ centres.

\begin{figure*}[htb]
\begin{center}
\includegraphics[width=120mm]{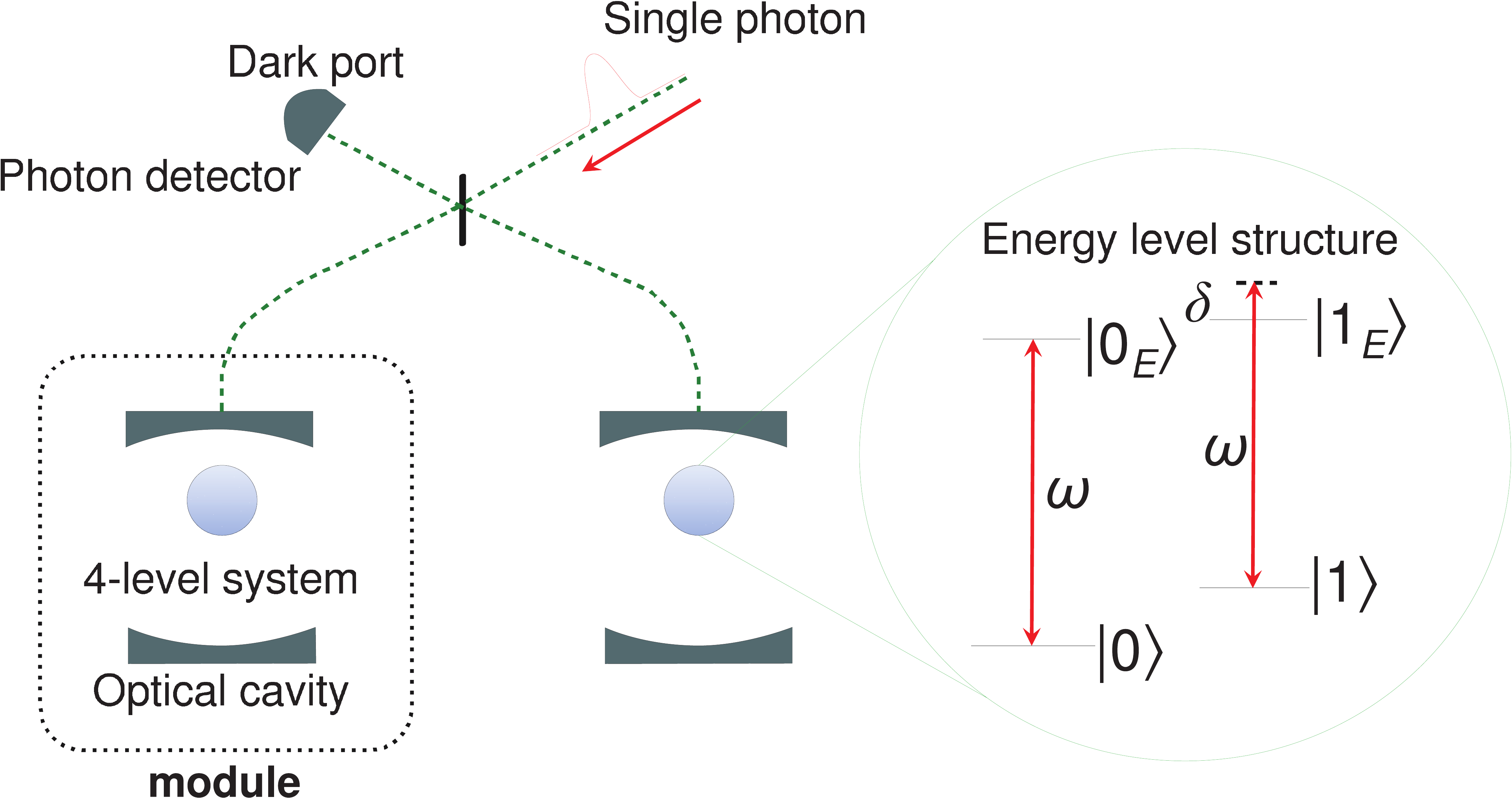}
\end{center}
\vspace{-10pt}
\caption{Schematic representation illustrating the module and the entanglement distribution scheme.  The module contains an optical cavity with a four-level system.  The entanglement distribution scheme is based on a Michelson interferometer where two modules are connected via an optical fiber. A single photon comes in from the right port and is conditionally reflected at each module depending on the state of the emitter.  Erasing the path information at the beam splitter followed by detection at the dark port projects the system to the singlet Bell state. } 
\label{fig1}
\end{figure*}

We begin our description of the architecture with an entanglement scheme based on the state-dependent reflectivity of a module consisting of an emitter-cavity system \cite{Waks2006,Hu2008}, as depicted in Fig.~\ref{fig1}. We can describe the emitter as a four-level system with transitions $|0\rangle \rightarrow |0_E\rangle$ and $|1\rangle \rightarrow |1_E\rangle$, each with a frequency $\omega_0$ and $\omega_1=\omega_0+\delta$, respectively. The probability for a photon to be reflected by a module with cooperativity $C$ and the cavity tuned to the interrogation frequency $\omega_0$ is given by \cite{Kimble1998}
\begin{equation}\label{eqn:reflection}
P_R=1-\frac{1+4 C+(\delta/\gamma)^2}{1+4 C+4 C^2+(\delta/\gamma)^2}.
\end{equation}
We have assumed a cavity with matched mirrors, in which case an impinging photon will be reflected by the module with high probability if the emitter is in the ground state $|0\rangle$ and the cooperativity is $C\gg 1$. In the case of large detuning, $(\delta/\gamma)^2\gg C^2\gg C\gg 1$, the cavity is effectively empty and the reflection probability approaches $P_R\rightarrow 0$. In the simplest variant of our entanglement scheme (Fig.~\ref{fig1}), we place two such modules at the output ports of a $50:50$ beamsplitter and prepare each emitter in an equal superposition of the ground states $|0\rangle$ and $|1\rangle$.  A single photon is then sent onto the beamsplitter. If it is subsequently detected at the ``dark" port of the beamsplitter, the emitters are projected onto the maximally entangled state
\begin{equation}
|S\rangle=\frac{1}{\sqrt{2}}(|1\rangle |0\rangle-|0\rangle|1\rangle )
\end{equation}
with success probability  $p=\eta^2/ 8$, where the collection efficiency $\eta^2$ includes the effects of inefficient sources and detectors and transmission losses. This probability may appear to be low, however  the generated entangled state has extremely high fidelity ($>$99\%) and is robust to imperfections (see supplementary material). For instance, imbalance in the cavity reflection coefficients slightly reduces the success probability but does not degrade the fidelity of the resulting state. 

The low success probability of the implementation can be simply overcome using a repeat until success approach to establish an entanglement link with high probability \cite{Barrett2005,Munro2010}. We will show in the following sections that the scheme not only exhibits high fidelity in the presence of physical imperfections, but also, unlike other approaches, does not involve any catastrophic errors. 

In addition, the module enables (near) perfect non-demolition measurement of the qubit state. For an architecture for quantum computation we require a second qubit in the cavity to act as a quantum memory.  Ideally, the coupling between our four-level system and this memory qubit can be switched on and off as required. This allows the four-level system to be reused for entanglement creation, now with a third module. By repeating this process with additional modules we can generate a cluster state suitable for fault-tolerant quantum computation. In the following we will detail this architecture by describing a full implementation using single NV centres in micro cavities connected in a photonic network.

\section{The diamond module} 

\begin{figure*}[htb]
\begin{center}
\includegraphics[width=140mm]{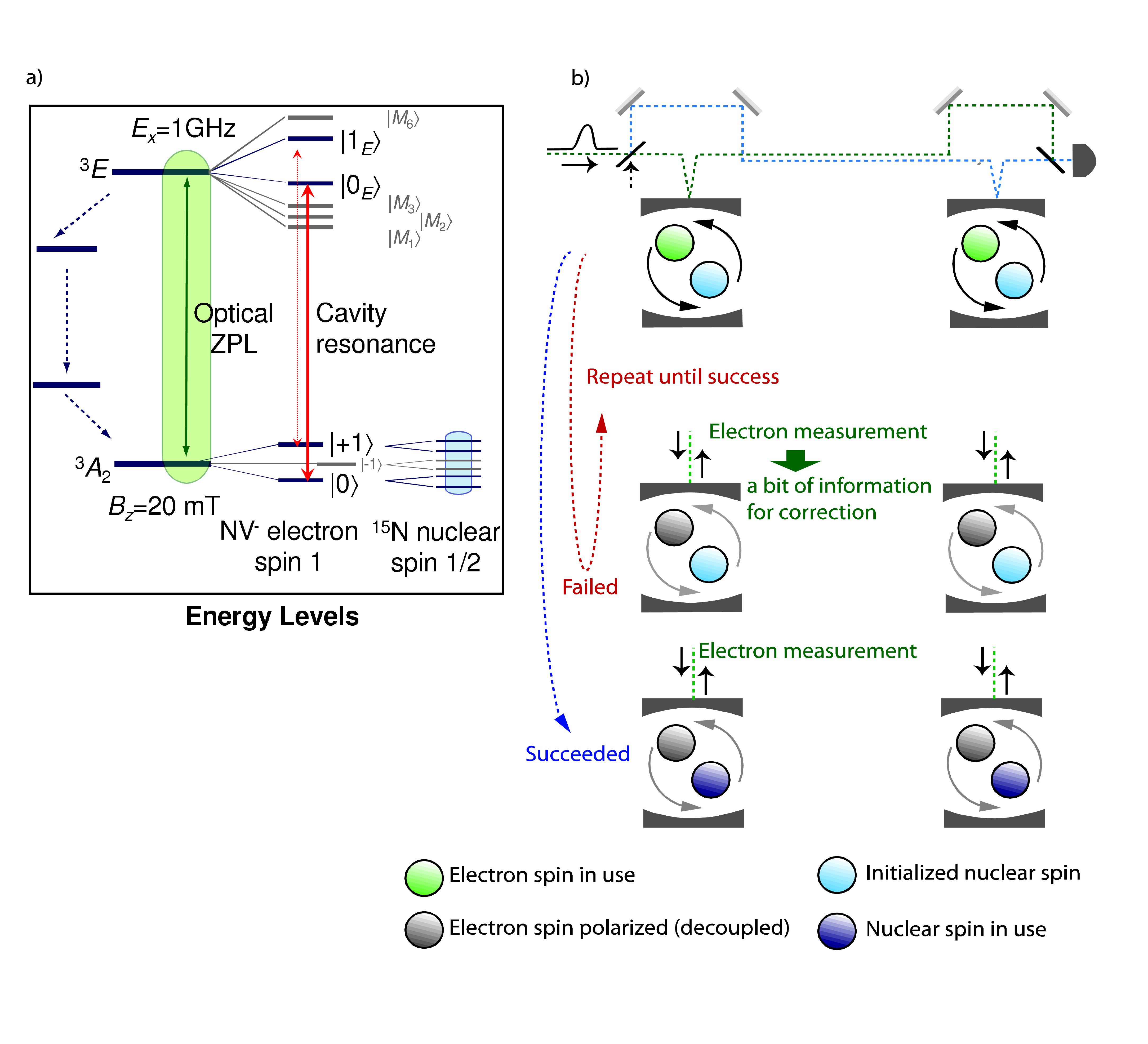}
\end{center}
\vspace{-10pt}
\caption{NV$^-$ centre is shown as a definite example of the artificial atom to realise the module.  Its energy level structure for a low temperature, low strain sample \cite{Tamarat2008,Togan2010} is illustrated in  a).  A static magnetic field of approximately 20 mT is used to separate the $m_s=\pm 1$ levels. The NV$^-$ centre possesses both an electron spin and $^{15}$N nuclear spin, which will be used to store and grow a cluster state for quantum information processing.  b) illustrates how the storage of entanglement in the nuclear spins is achieved. The nuclear spin needs to be prepared in the superposition state, $|n_+\rangle= \frac{1}{\sqrt 2}(|0\rangle +|1\rangle)$ before the protocol starts.  During this operation, the electron spin is in a polarized state $|0\rangle$, hence the hyperfine coupling is effectively turned off.  When the electron spins rotate to $\frac{1}{\sqrt 2}(|0\rangle +|+1\rangle)$ for the entanglement distribution scheme, the clock associated with the hyperfine coupling starts. A spin-echo sequence can be used to decouple the electron and nuclear spins where necessary---for instance, when the  entanglement distribution fails, we need to decouple the electron and nuclear spins before re-initialising the electron spin and attempting the protocol again until success. } 
\label{fig2}
\end{figure*}

Let us now turn our attention to a concrete implementation:  a fiber-connected optical cavity containing a single NV$^-$ centre, of which the energy levels are depicted in Fig.~\ref{fig2}a. 
The  lowest three electron spin states, $|m_s=0,\pm 1\rangle\equiv |0,\pm 1\rangle$ form the spin-1 $^3A_2$ manifold which has a zero-field splitting of $2.87$ GHz.  With an externally applied magnetic field $B\sim 20$ mT, our electron spin qubit levels $|0\rangle$ and $|+1\rangle$ are far detuned from the $|-1\rangle$ energy level and so form an excellent qubit.  The isotope $^{15}$N will be utilised as a spin-\nicefrac{1}{2} nuclear memory. Next, the optical transitions between one of the $^3A_2$ magnetic sub-levels $|i\rangle$ and the $^3E$ levels $|M_i\rangle$  coupled to the cavity field can be represented by $\hbar g_{m_s,i} \sum_{i=1...6}  \left[ a^\dagger |i \rangle \langle M_i |  + a | M_i\rangle \langle i | \right]$ where  $g_{m_s,i}$ are the coupling constants between the transitions and field, $a^\dagger, (a)$ are the field's creation (annihilation) operators and $M_i$ are the energy eigenstates, in order of ascending energy, within the $^3E$ manifold. At zero strain they are given by the basis states $\{M_{1...6}\}=\{E_2,E_1,E_x,E_y,A_1,A_2\}$, neglecting a small mixture of the $E_{x,y}$ and $E_{1,2}$ states due to spin-spin interaction. The basis states $E_x$ and $E_y$ have electronic spin zero, while the others ($A_{1,2}$ and $E_{1,2}$) are equal superpositions of spin $\pm 1$ \cite{Togan2010,Maze2011}. For our scheme, we apply an electric field in the $x$-direction (${\cal E}_x$) to lift the degeneracy of the spin-zero states in the excited-state manifold.  This greatly reduces the sensitivity to rogue strain or electric field influences in the $y-$direction and thus makes the system more robust. ${\cal E}_x$ can be adjusted at each site to bring different NV$^-$ centres to the same resonance frequency. We choose $| 0_E\rangle=| E_x\rangle+\epsilon$ and $| 1_E\rangle| =| M_5\rangle=0.98| A_1\rangle+0.17| A_2\rangle+\epsilon$, where $\epsilon$ represents negligible contributions from other basis states. For this setting, we find $\delta=2\pi\times 2.71\,$GHz, which is far greater than the homogeneous optical half-width of the chosen transitions, $\gamma=2\pi\times 11\,$MHz. We note that although the NV$^-$ is not a simple four-level system (Fig.~\ref{fig2}a), all other allowed transitions are detuned even further from the excitation frequency $\omega$ and can be neglected. Thus we have the properties required for entanglement distribution based on state-selective reflection using the NV$^-$ centre electron spin states $|0\rangle$ and $|+1\rangle$. 

\subsection{Quantum non-demolition detection}

The conditional reflection of a photon from a module allows us to perform a quantum non-demolition measurement of the NV$^-$ state \cite{Volz2011} (see supplementary material). The measurement sequence consists of a photon measurement followed by a qubit flip and a second photon measurement. For the photon measurement, a single photon is sent to a module, and will be reflected and detected if the NV$^-$ centre is in the state $|0\rangle$, and lost otherwise. The qubit flip is achieved by a microwave $\pi-$pulse. A photon detection would be expected with certainty for one of the photon measurements under ideal conditions, while the absence of a detection event would indicate leakage of the NV$^-$ centre from the qubit subspace to the $|-1\rangle$ state. The destructiveness of this measurement depends on the probability of exciting the NV$^-$ centre and the subsequent spin-flip probability. The measurement needs to undergo several repetitions to make up for finite photon collection efficiency, thereby increasing the spin-flip probability. Nonetheless, we find that it is possible to achieve a measurement error rate of $\epsilon_{QND}=0.1\%$ even for a finite collection efficiency of $\eta^2=0.3$, which is sufficient for fault-tolerant computation (see Section \ref{sec:benchmarking}). For unity detection efficiency, $\epsilon_{QND}$ reaches $0.01\%$, but cannot be reduced further in our current scheme due to the non-zero spin-flip probability for each measurement. This is due in part to off-resonant excitations of the NV$^-$ centre in the cavity, the probability of which increases with cooperativity.  This leads to an working cooperativity of $C\simeq 50$, which is realistically achievable with currently available microcavity technology. 

\subsection{Remote entanglement}
We begin by initialising each electron spin to $|0\rangle$ followed by rotating to $\frac{1}{\sqrt{2}}(|0\rangle+|+1\rangle)$ using a polarised driving field in a few nanoseconds. A single-photon pulse is then sent onto the interferometer  (Fig.~\ref{fig2}b) and the dark port monitored. We repeat this procedure until the entanglement is heralded by the successful detection of a photon at the dark port.  This is made possible by the good cycling properties of the NV$^-$ transition $|0\rangle \rightarrow |E_x\rangle$ \cite{Togan2010}. We note furthermore that the de-ionization process NV$^-\rightarrow$ NV$^0$, and the resulting dynamical spectral diffusion, is rendered impossible by using only one single-photon excitation in the interferometer at a time \cite{Togan2010}.

In the NV implementation of our module, the nuclear spin is a long-lived quantum memory which will in our architecture be designated to store one node of a cluster state \cite{RB2001}. Our scheme creates entanglement between the electrons of the two NV$^-$ centres.   The transfer of the entanglement to the nuclear spin memories is done through the Ising component of the hyperfine coupling ($A_\parallel \sim 3.03$ MHz \cite{Felton2009}), which is tuned by the external magnetic field of  $B\sim 20$ mT to give a conditional phase on the state of the two spins. The amount of entanglement oscillates in time from zero to maximum.  At time $\tau$ setting $\pi$ points of the oscillation, the effective gate becomes a controlled-phase gate, while at the $2\pi$ point it gives identity.  The hyperfine coupling is always present but is effectively turned off while the electron spin is in the polarised state $|0\rangle$.

Putting this together, the complete nuclear spin entanglement protocol begins with both electron spins and both nuclear spins polarised in their ground states (achieved via the quantum non-demolition measurement). The electron spin is then rapidly rotated to the $|+\rangle$ state via a $\pi/4$ $Y$ rotation, an effective controlled-NOT operation is then performed between the nucleus and electron via the hyperfine interaction at which point the electron is again rotated by  $\pi/4$ around the $Y$-axis and measured in the computational basis.  This initialises the nuclear spin into the $\ket{n_+}$ state.  We then rotate the electron back into the $\ket{+}$ state to attempt an electron-electron bond via the optical transitions.  The hyperfine coupling turns on when the photonic entangling protocol is initiated by the electron spin rotation but  a spin echo like sequence can be used to disentangle the electron and nuclear spins at any time we require. If the gate has succeeded, we perform a $\pi/4$ $Y$-rotation on one of two electron spins, and wait until the hyperfine interaction maximally entangles the electron and nuclear spins within each node.  A $\pi/4$ $Y$-rotation is then performed on the electron spin of each module followed by its measurement in the computational basis. This completes the transfer of the entangled link to the nuclear spins.  If the entanglement distribution has failed, the protocol will be repeated until a success is heralded, as illustrated in Fig.~\ref{fig2}b. We note that it is not necessary to reinitialise the nuclear spin prior to each attempt.

\section{Sharing entangled states between three modules}  

\begin{figure*}[htb]
\begin{center}
\includegraphics[scale=0.6]{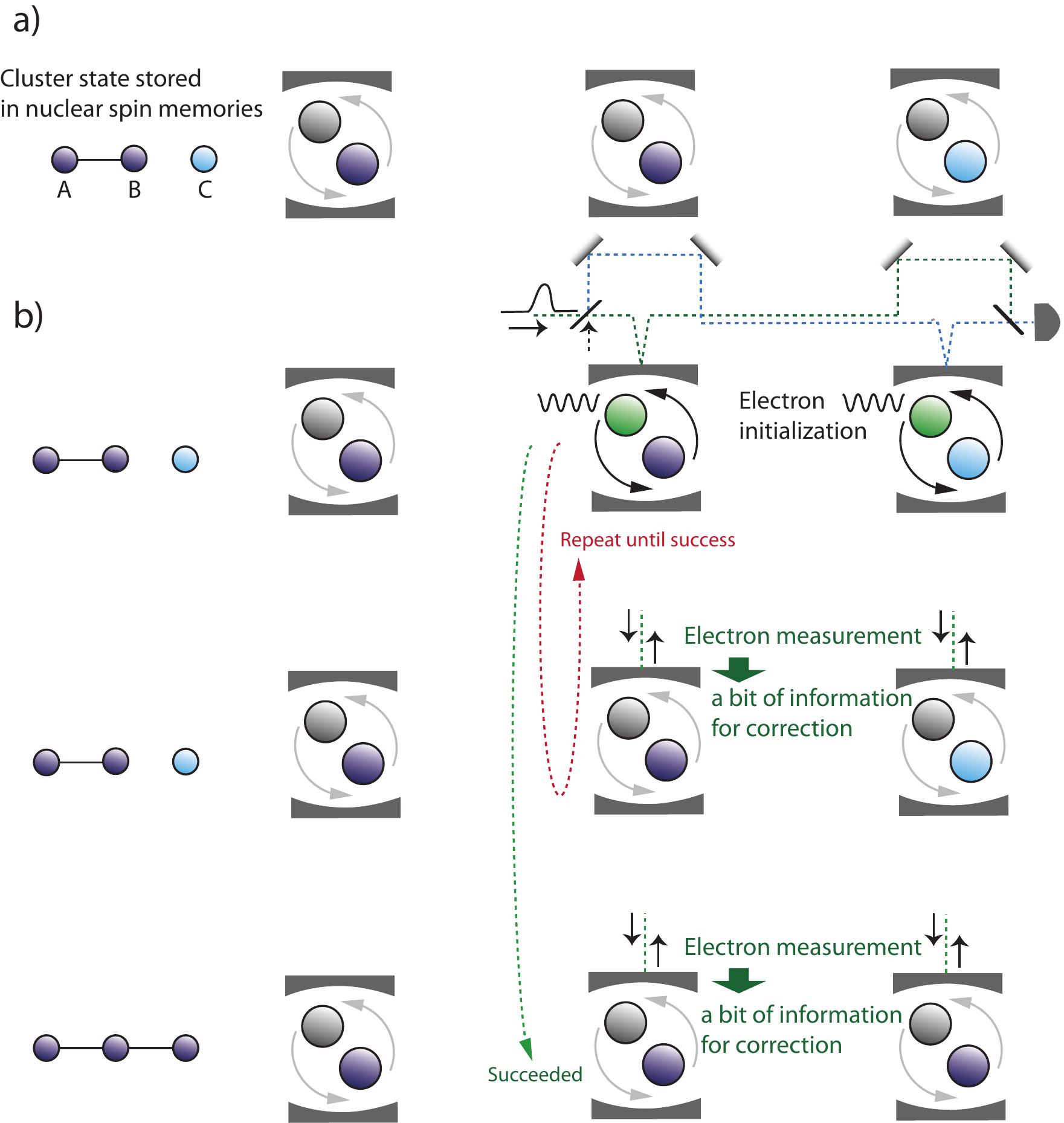}
\end{center}
\vspace{-10pt}
\caption{The repeat until success protocol is accurately time sequenced.  This is required by the nature of the coupling, as entanglement between the electron spin and nuclear spin oscillates.  Upon failure, we wait until the $2\pi$ point in the entangling cycle, where the nucleus and electron are decoupled.  The nuclear spin is consequently protected from  feedback errors through the hyperfine coupling by accurately timing the re-initialisation of the electron spins.  When the distribution of entanglement between two electrons succeeds, the entanglement bond will be transferred to the nuclear spins by waiting until a $\pi$ point where the electron and nuclear spins are maximally entangled.} 
\label{fig3}
\end{figure*}

The next step is to extend our cluster of two nuclear spin qubits to three (by adding one). We begin with an entangled pair stored in the nuclear spins of modules A and B as shown in Fig.~\ref{fig3}. A new entanglement bond on the electron spins in modules B and C is created using the same repeat until success protocol, though only the nuclear spin in C will be initialised.  Once the entanglement between the electronic qubits is created, the entanglement will be transferred to the nuclear spins using the hyperfine coupling described previously.  

This time, the nuclear spin in module B is in use, carrying information established at the beginning of the protocol. Photon loss may feedback via the permanent hyperfine coupling, introducing catastrophic errors in the states stored in the nuclear spins in modules A and B. For the protocol to be useful, we should be able to preserve with high-fidelity the existing entangled states stored in the nuclear spins of A and B,  while using the electron spin in B to create new entanglement with module C.

By introducing a time-sequenced entangling procedure we can avoid decoherence caused by photon loss. Furthermore, by using spin-echo like sequences to decouple the electron spins from their surrounding environment we may extend their coherence time. The clock for the hyperfine coupling sequence starts when the photonic entangling protocol  is initiated (that is, when the electron spin is rotated out of a polarised $|0\rangle$ state). If the entangling protocol fails, the system waits until the spin echo sequence decouples the electron and nuclear spins. At this point the nuclear system recovers coherence and the information stored on the nuclear spin remains untouched until the protocol succeeds. This process is illustrated in  Fig.~\ref{fig3}.  Once the new entangling bond is established, indicated by a heralding signal, we again wait until the spin echo decouples the electron and nuclear spins. We then perform a single $\pi/4$ $Y$-rotation on one of the two electron spins, and wait until the hyperfine interaction maximally entangles the electron and nuclear spins within each the nodes.  An $X$-basis measurement is performed on each electron (via a $\pi/4$ $Y$-rotation and computational basis readout) to transfer the new bond to the nuclear system. 
  
\begin{figure*}[htb]
\begin{center}
\includegraphics[scale=0.55]{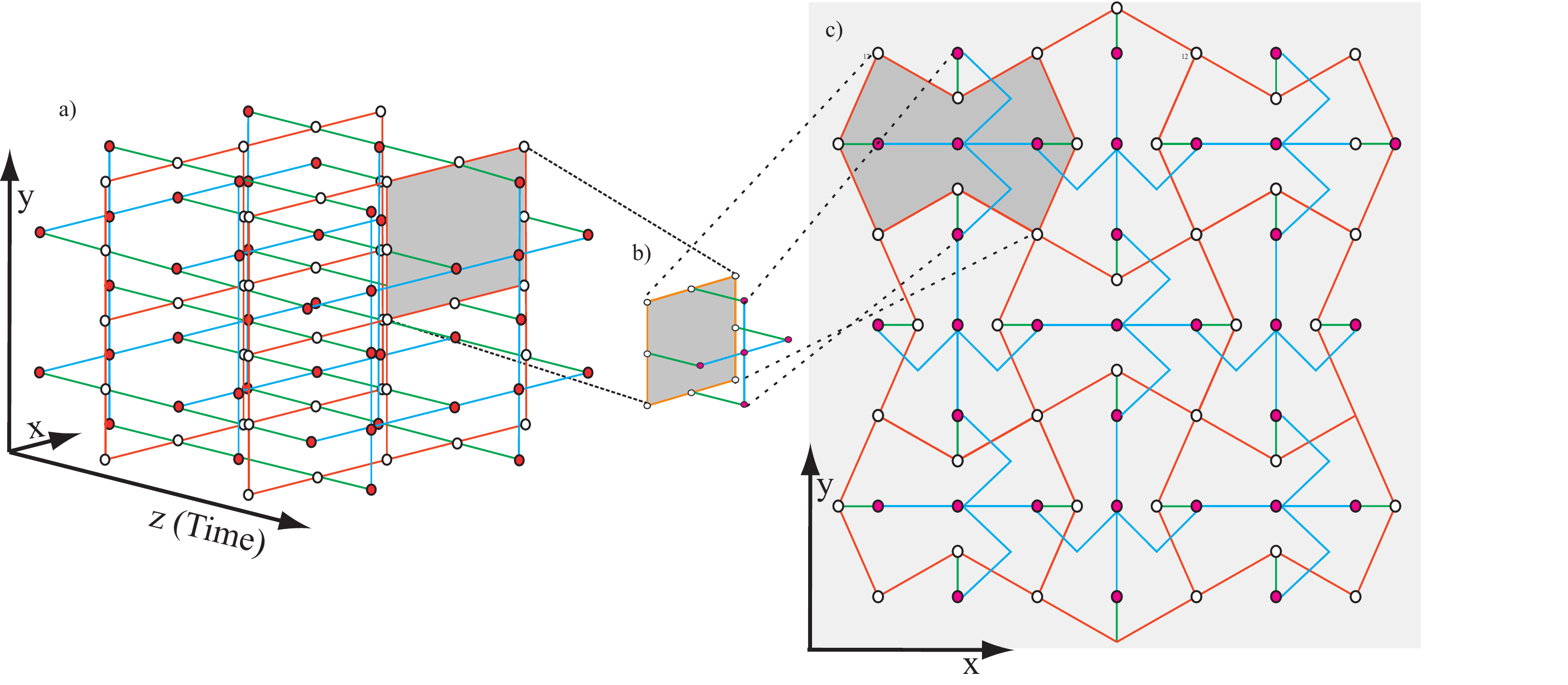}
\end{center}
\vspace{-10pt}
\caption{Three-dimensional topological cluster state and module connectivity in a two-dimensional plane.  (a) The topological cluster state cluster is a resource for fault-tolerant quantum computation. However, the whole state is not required at all times during the computation. Instead, only two layers of the cluster state need to be prepared and stored at any given time. b) The physical unit cell composed of two layers. The back layer contains eight connected qubits arranged in a square (orange), while the front layer has five qubits arranged in a cross (blue). The two layers are connected by controlled-phase gates (green). Measurement of the front layer of the cluster will teleport the current state of the computer to the back layer, at which point the physical qubits we just measured can be reconnected in accordance with the geometry of the cluster state, and the information can be teleported back again. In this way, the two physical layers execute the even and odd temporal steps of the computation, allowing an arbitrarily deep computation to be performed with a fixed number of physical qubits. c) A compact layout of modules on two-dimensional plane. } 
\label{fig4}
\end{figure*}

Repeating this with additional modules we can generate an arbitrary cluster state. We are particularly interested in generating the three-dimensional topological cluster state (illustrated in Fig.~\ref{fig4}a) capable of supporting fault-tolerant quantum computation \cite{RHG07,FG09}. Topological models of error correction \cite{Kitaev2003,Dennis2002} exhibit relatively high tolerance to errors and are particularly well suited to architectures due to their simple underlying structure \cite{DFSG08,SJ09,JMFMKLY10,YJG12,Nickerson2013}. The topological cluster state is particularly useful in the context of our repeat until success protocol as it is inherently robust against missing bonds, which will be heralded. These missing bonds can be processed in the classical interpretation of measurement results, without any modification to the quantum circuit \cite{Barrett2010}. To prepare the topological cluster state, each physical qubit is entangled with its four nearest neighbours, hence a dagger shaped cluster state is the fundamental unit, independent of the size of the network, highlighted by blue bond in Fig.~\ref{fig4}b. Four entangling steps are required to create this fundamental state with five modules.

\section{Benchmarking the photonic architecture} 
\label{sec:benchmarking}

To process quantum information with a three-dimensional topological cluster state, the state is consumed by measurements on physical qubits in sequential two-dimensional layers, where one axis is defined as the temporal axis. These measurements create and manipulate encoded qubits defined by defects \cite{RHG07,Devitt2012}. As the computation proceeds by measuring one layer at a time, the whole topological cluster state is not required to be constructed initially. Only two successive layers need to be prepared and stored at any given time, allowing us to concentrate on only two physical layers of modules. The current state of the computer is teleported back and forth between these two layers, which are refreshed and recycled to generate the entire topological cluster. Taking the centre of each cell (in Fig.~\ref{fig4}b), we initiate a sequence of gates to generate the dagger shaped cluster state throughout the lattice, which generates one layer of the topological cluster state (see supplemental material). The two layers of the module network are flattened to a two-dimensional plane, as shown in Fig.~\ref{fig4}c. This pattern repeats to an arbitrarily large cross section.

\begin{figure*}[htb]
\begin{center}
\includegraphics[scale=0.75]{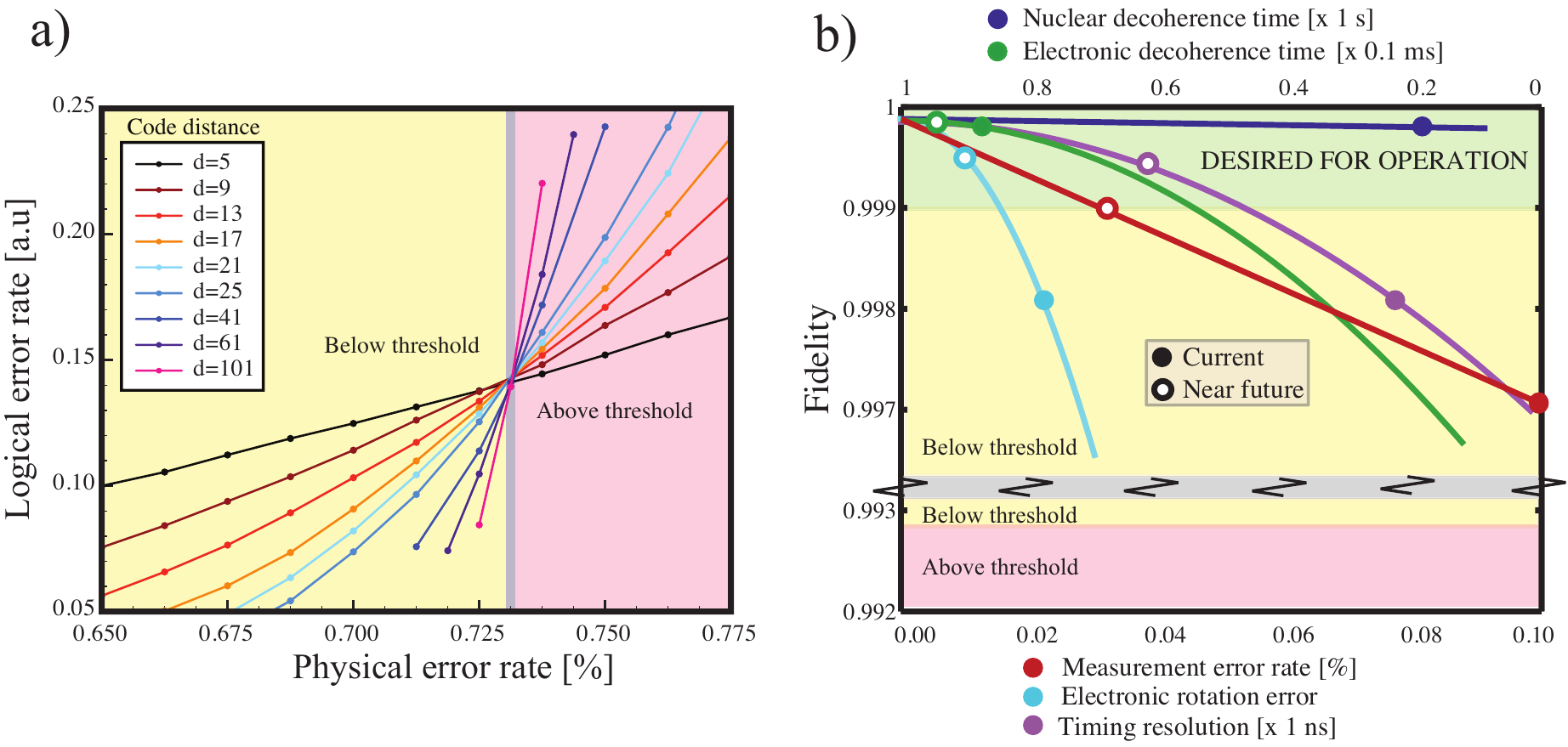}
\end{center}
\vspace{-10pt}
\caption{Fault-tolerant thresholds and required component error rates.  a) Numerical simulation of topological error correction. The logical error rate is plotted as a function of the physical error rate for various code sizes (distances d), where we have assumed that all gates and measurements are operating at the same error rate. Each point corresponds to at least $10^4$ trials. The value of the physical error rate at the intersection gives the threshold (in this case, approximately $7.3\times10^{-3}$). For physical error rates below this threshold, the logical error rate can be reduced arbitrarily by increasing the code distance. b) The required fidelity for each physical parameter. The dots on the lines show the current best accuracy reported, all of which already meet the required accuracy, 99.27\%. For a realistic implementation, the gate fidelity should be above 99.9\%, corresponding to the green coloured regime in the plot. Electronic and nuclear spin coherence times are already in this regime, and the remaining parameters may soon meet the desired accuracy given the rapid development of quantum control of such systems. The fidelity does not converge to unity due to imperfections in the NV$^-$ centre.} 
\label{fig5}
\end{figure*}

At a circuit level, we are interested in the threshold error rate, below which the architecture becomes fault tolerant \cite{RHG07}. The projective measurement of  the nucleus (via the electron-nucleus hyperfine interaction) allows us to combine measurement and reinitialisation of the nuclear qubit in a single step. Therefore, the depth of the quantum circuit to prepare the topological cluster state is reduced from six steps to five. We find that this reduction increases the error threshold to 0.73\% (see Fig.~\ref{fig5}a). Given this threshold, our target error rate for the five relevant gates is $\sim 0.1$\%, as this is sufficiently far below the threshold to allow significant suppression of errors using a practical number of modules \cite{Devitt2012}.

The target error rate does not tell us much until it is decomposed into each physical component. Each gate consists of several physical steps and involves several sources of errors. In our case, these sources are parameterised by the nuclear and electron spin decoherence times, electron measurement efficiency, electron rotation efficiency, and timing error. As described, the sequence to generate an entangling bond is probabilistic, and the protocol repeats until success. Given that we require bonds to succeed with probability $P=99.9$\%, if the success probably of a single attempt is $p_c$, the number of attempts we require is $s = log(1-P)/ log(1- p_c)$. For $p_c = 0.0625$, $s = 107$. We consider the error rate for each gate to be the worst-case scenario, as heralded failure can be significantly higher than the error rate for unheralded errors \cite{Barrett2010}.

The required fidelity for each physical parameter is shown in Fig.~\ref{fig5}b. Each curve is plotted assuming it is the \emph{only} non-zero error, except for possible errors arising from the absorption of photons by the NV$^-$ node (see supplementary material). The coloured green region in Fig.~\ref{fig5}b is the target for each parameter for an operational computer (though parameters in the yellow region still lead to gates below the threshold).  For the architecture to be fault tolerant, these errors need to be combined (see supplementary material).  Electron and nuclear decoherence is already sufficiently low \cite{Balasubramanian2009,Ishikawa2012,Fang2013,Maurer2012}, while the other parameters still need improvement. However, it is important to note that the required improvements are less than one order of magnitude, and are not limited by any currently known fundamental limitations of the NV$^-$ system itself. 

Assuming that the threshold condition is met, performance is mostly dependent on the computational cycle time, which is limited by the time taken to establish all the electron-electron connections.  For bond connections with $P=99.9$\%, the total time required to create a nuclear-nuclear bond is 3.5 $\mu s$, assuming $p_c=$ 6.25\%. This time could be reduced by lowering the required connection efficiency and exploiting the robustness of the topological code to missing bonds \cite{Barrett2010}. The quantum circuit takes five steps to construct each cross-sectional layer of the topological cluster state.  Hence, a unit cell of the cluster is prepared every $\sim 30\;\mu$s.  To implement an algorithm on the computer, we create pairs of defects in the cluster.  The volume of cluster allocated to pairs of defects represents the degree of error correction, parametrised by the distance between defects, $d$. For a logical error rate $p_L \leq 10^{-18}$, $d \geq 32$ is required \cite{Devitt2012}. Therefore, a logical cell requires $V=\left(\frac{5d}{4}\right)^3 = 40^3$ cluster cells.  To perform a logical CNOT gate requires a cluster volume $2\times 2$ in cross section and 2 logical cells in temporal depth.  Hence, it takes $3.4$ ms for $p_c=$ 6.25\% (a clock frequency of $\sim$ 295 Hz).  This rate can be further improved by better optical efficiencies, but is ultimately limited by the hyperfine interaction of the  NV$^-$ node used for nuclear spin operations.  If we assume a deterministic electron-electron connection, a logical CNOT gate would take approximately 960 $\mu$s ($\sim$1 kHz) as the system becomes rate limited by nuclear measurement  (see supplementary material).

\section{Discussion} 
As we have seen, a simple module can form the basis of a scalable quantum computer architecture. The architecture is naturally distributed, and hence is applicable to quantum communication \cite{Stephens2012}. Such a network may be local or global, with local networks connected by quantum communication channels. In this case, the distance between the modules may become orders of magnitude larger. The time delay due to the communication distance may be mitigated by the long-lived memory inside the module. With increased distance between modules, photon loss would increase, reducing the success probability of the entangling protocol. However, long-distance communication does not necessary require $P=99.9$\%. Instead, with $P=99.0$\%, the number of attempts can be reduced to $s=71$ for $p_c=$ 6.25\%.

We have found that physical requirements of our architecture are broadly consistent with present technology. However, improvements are still required, in particular to the measurement efficiency. However, while technological developments might help to meet these requirements, physical requirements may be found to be less stringent with a more sophisticated adaptive error analysis. 

\vskip 1 truecm
\emph{Acknowledgements---} We thank Austin Fowler, Andrew Greentree and Burkhard Scharfenberger for valuable discussions. We acknowledge partial support from FIRST, NTT and NICT  in Japan, the Austrian Science Fund (FWF) through the Wittgenstein Prize and the EU through the project DIAMANT.  KB and TN acknowledge support from the FWF Doctoral Programme CoQuS (\textit{W1210}).

\appendix
\section{Description of the NV$^-$ centre}
The dynamics of the NV$^-$ centre, consisting of the electron spin-1 $^3A_2$ manifold and the nuclear spin-1/2 system, can be described by the Hamiltonian $H= H_{\rm e}+H_{\rm n}+H_{\rm e-n}$. The electron spin's ground state Hamiltonian is given by \cite{He1993,Manson2006}
$$H_{\rm e}  = \hbar ( D S_z^2 +E \left[S_x^2 - S_y^2\right]  + g_e \mu_B B S_z),$$
which represents a zero-field splitting ($D/2 \pi = 2.87$ GHz), a strain induced splitting ($E/ 2 \pi \sim 1$-$10$ MHz), and a magnetic field induced splitting ($g_e \mu_B B$), where $\mu_B$ is the Bohr magneton and $g_e=2.0$ is the g-factor. In this Hamiltonian, $S_z, S_x,S_y$ are the usual spin-1 operators. With an externally applied magnetic field $B\sim 20$mT, our electron spin qubit levels $|0\rangle$ and $|+1\rangle$ are far detuned from the $|-1\rangle$ energy level, supporting our electron spin qubit. The nuclear spin Hamiltonian $H_{\rm n} = -\hbar    g_n \mu_n B I_z$ represents a magnetic field induced splitting of the $^{15}$N nuclear spin, where $\mu_n$ is the nuclear magneton and $g_n=-0.566$ the nuclear g-factor. $I_z$ is the usual Pauli $Z$ spin-1/2 operator. 

The hyperfine coupling between the electron and the nuclear spins is given by \cite{Felton2009}
$$H_{\rm e-n} =  \hbar   A_{\parallel} S_z I_z+ \frac{\hbar  A_{\perp}}{2} \left(S_+ I_- + S_- I_+ \right),$$
 where $S_\pm$ ($I_\pm$) are the electron spin (nuclear spin) raising and lowering operators respectively. This coupling includes an Ising part with coupling strength $A_{\parallel}/2 \pi \sim 3.03$ MHz and an exchange part with coupling constant $A_{\perp}/2 \pi \sim 3.65$ MHz \cite{Felton2009}. With $B\sim 20$ mT the exchange coupling is far off-resonance resulting only in a small dispersive phase shift.   This results in a natural controlled-phase gate that operates on a time scale $\tau \sim \pi/  \left[ A_{\parallel}+\frac{A_{\perp}^2}{2 \lambda}\right] \sim 165$ ns, where $\lambda$ is the frequency difference between the electron and nuclear spin levels.  

An external microwave driving of amplitude $ \Omega_0$ is used to perform the electron and nuclear spin rotations.  The driving Hamiltonian can be expressed as 
$$H_{\rm d}= \hbar \Omega_0 \cos \left(\omega_d t + \phi \right) \left( S_x - \frac{g_n \mu_n}{ g_e \mu_B} I_x\right),$$
where the frequency $\omega_d$ is chosen appropriately to determine whether we drive the electron or nuclear spin, with $\phi $ representing an initial phase offset. By using a polarised field, electron spin rotations can be achieved with high fidelity in at most a few nanoseconds. The nuclear spin operations are much slower due to the weak gyromagnetic ratio but can be achieved (with high fidelity) in a few microseconds by using the hyperfine coupling to enhance the natural nuclear spin splitting. 

Next, the NV$^-$ centre also has a $^3E$ energy level manifold with optical transitions to the $^3A_2$ manifold. The optical transitions between one of the $^3A_2$ magnetic sub-levels and the $^3E$ levels coupled to the cavity field can be represented by 
$$H_{\rm e-f} = \hbar g_{m_s,i} \sum_{i=1...6}  \left[ a^\dagger |i \rangle \langle M_i |  + a | M_i\rangle \langle i | \right],$$
where $M_i$ are the energy eigenstates, in order of ascending energy, within the $^3E$ manifold. At zero strain and magnetic field, the $^3E$ manifold is represented by the basis states $\{M_{1...6}\}=\{E_2,E_1,E_x,E_y,A_1,A_2\}$, neglecting a small mixture of the $E_{x,y}$ and $E_{1,2}$ states due to spin-spin interaction.  The optical field of frequency $\omega$ can be described by $H_{\rm f} = \hbar \omega a^\dagger a$ with $a^\dagger$ $(a)$ being the field's creation (annihilation) operators.   The cavity coupling rate for a given transition is given by $g_{m_s,i}$. The basis states $E_x$ and $E_y$ have electronic spin zero, while the others ($A_{1,2}$ and $E_{1,2}$) are equal superpositions of spin $\pm 1$ \cite{Togan2010,Maze2011}. For our scheme, we apply an electric field of $E_x=1\,$GHz in the $x$-direction so as to lift the degeneracy of the spin-zero states in the excited-state manifold.  This greatly reduces the sensitivity to rogue strain or electric field in the $y-$direction making the system more robust. 

\subsection{Coherence Properties}
It is critical to mention the coherence properties of our electron-nuclear spin as this can vary significantly. Here we are assuming a single $^{15}$NV$^-$ centre is created on isotopically pure (99.9\%+ 12C) diamond substrate \cite{Balasubramanian2009} and that our module will operate at low temperature (4-20 K) rather than room temperature. In such a case it has been reported that $T_1$ of the electron spin is greater than 1 s, while $T_2^* \sim 90$ $\mu$s \cite{Ishikawa2012,Fang2013} with $T_2$ much longer \cite{Balasubramanian2009}. The nuclear spin $T_1$ and $T_2$ are at least 0.2 s at present \cite{Maurer2012}. The limiting coherence parameter in this design is the $T_2^*$ of the electron spin during the 165 ns controlled-phase gate. However with Gaussian decay having the form $\exp \left[ - \left(2 t / T_2^*\right)^2\right]$, the error associated with this is small in principle ($<10^{-5}$) \cite{Balasubramanian2009}.

\section{The diamond module}

\begin{figure}
\begin{center}
\includegraphics[scale=1.2]{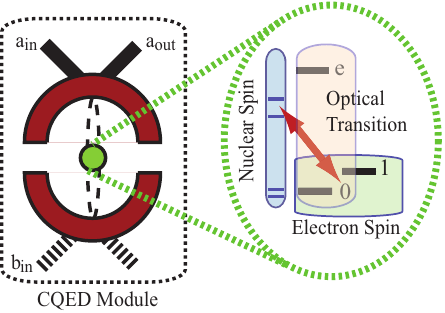}
\end{center}
\vspace{-10pt}
\caption{Schematic representation of a repeater node containing an optical cavity with an embedded NV$^-$ centre.  The NV$^-$ centre possesses both an electron spin and a $^{15}$N nuclear spin. A static magnetic field of approximately 50 mT is used to separate the $m_s=\pm 1$ levels. } 
\label{main-fig}
\end{figure}

At the centre of our approach is a quantum module in which an NV$^-$ centre is embedded in an optical cavity (Fig.~\ref{main-fig}). The NV$^-$ centre is composed of a spin-one electronic spin and a spin-half $^{15}$N nuclear spin. Our module is an interface between the optical, microwave and radio frequency regimes allowing information to be transferred between them. It works as follows: state dependent reflection allows the creation of entanglement between an external optical field \cite{Waks2006,Hu2008,Young2009,Santori2010} and the electron spin, while the hyperfine interaction allows the transfer of  the electron spin state to the long-lived nuclear spin. It also allows the nuclear spin to be measured via the electron spin, thus completing the interface. While this is conceptually simple, the details of the physical system lead to a number of complications which we will address in this supplementary material. 

To understand exactly how this module operates we must examine the interactions between the three components of our hybrid system (optical field, electron spin, nuclear spin) as a whole. The overall system  including couplings and driving fields can be described by the Hamiltonian
\begin{eqnarray}
H= H_{\rm f}&+& H_{\rm e}+H_{\rm n}+H_{\rm d} +H_{\rm e-n}+H_{\rm e-f},
\end{eqnarray}
where $H_{\rm f} = \hbar \omega a^\dagger a$ is the Hamiltonian for the optical field detuned from the cavity resonance frequency $\omega_c$ by $\Delta=\omega_c-\omega$ with $a$ $ (a^\dagger)$ being the field annihilation (creation) operator. 

The second term $H_{\rm e}  = \hbar ( D S_z^2 + E \left[S_x^2 - S_y^2\right]  + g_e \mu_B B S_z)$ represents a zero field splitting ($D/2 \pi = 2.87$ GHz), a strain induced splitting ($E/2 \pi < 10\,$MHz) and a magnetic field induced splitting ($g_e \mu_B B$)  for the NV$^-$ centre's electron spin  \cite{Felton2009}. In this spin-one system,  $S_{x,y,z}$ represents  the generalised Pauli $X$,$Y$,$Z$ operators with $S_+$ ($S_-$) being the raising (lower) operator. Further $\mu_B$ is the Bohr magneton and $g_e=2.0$ the g-factor. For an externally applied magnetic field  of $B\sim 20$ mT, the $|0\rangle$ and $|+1\rangle$ levels are separated by approximately $3.43$ GHz. The $|m_s=-1\rangle$ energy level is detuned approximately $1.1$ GHz below the  $|m_s=+1\rangle$ level and $\sim2.3$ GHz above the $|m_s=0\rangle$ level. 

The third term $H_{\rm n} = -\hbar    g_n \mu_n B I_z$ represents a magnetic field induced splitting of the  nuclear spin with $I_z$ being the Pauli $Z$ spin-half operator. Here, $\mu_n$ is the nuclear magneton and $g_n=-0.566$ the nuclear g-factor. The computational basis states of the nuclear spin are  $|\downarrow\rangle$ ($|\uparrow\rangle$). 

Next, $H_{\rm d} = \hbar \Omega_0 \cos \left(\omega_d t + \phi \right) \left( S_x - \frac{g_n \mu_n}{ g_e \mu_B} I_x\right)$ represents an electromagnetic field  driving whose magnitude on the electron (nuclei) is determined by both the amplitude $\Omega_0$ of the applied field and the ratio of $g_n \mu_n / g_e \mu_B$. The  frequency $\omega_d$ is chosen appropriately to determine whether we drive the electron or nuclear spin while $\phi $ specifies the phase. 

The first of the coupling terms $H_{\rm e-n} =  \hbar   A_{\parallel} S_z I_z+ \hbar \frac{A_{\perp}}{2} \left(S_+ I_- + S_- I_+ \right)$ represents a hyperfine interaction between the electron and nuclear spin. This coupling contains both an Ising part with coupling strength $A_{\parallel}$ and an exchange part with coupling constant $A_{\perp}$. For a $^{15}$N nucleus, $A_{\parallel}/2 \pi \sim 3.03$ MHz and $A_{\perp}/2 \pi \sim 3.65$ MHz \cite{Felton2009}.  The second coupling term $H_{\rm e-f}$ is between the optical field and the electronic spin. It is detailed in the methods section of the main text and will be discussed in the next several sections. 

Before proceeding it is also useful to consider the coherence parameters of  our NV$^-$ centre. With isotopically purified CVD diamond \cite{Maze2008,Ishikawa2012,Fang2013} we can expect electronic spin coherence times $T_2^*$ of 90 $\mu$s and  $T_2 > 1.8$ ms while the relaxation $T_1$ can be over 1 second when the sample operates in the 4-80 K regime \cite{Jarmola2012}. The coherence times of nuclear spins have been shown to exceed 1 s \cite{Maurer2012}. We now explore in detail measurement and entanglement of two NV$^-$ centres.

\subsection{Level structure}

In this section we consider the main features of an NV$^-$ centre in a microcavity to ascertain how well the state of an NV$^-$ centre can be coupled to an external optical field and detected, and how two NV$^-$ centres can be entangled by detection. We apply a magnetic field $B_z=20\,$ mT to separate the ground state levels $\vert \text{+1}\rangle$ and $\vert-1\rangle$. We aim to use resonant light tuned to the $\vert 0\rangle \leftrightarrow \vert M_{3}\rangle\equiv(0.998\vert E_{y}\rangle+0.07 \vert E_{1}\rangle)\simeq \vert E_{y}\rangle$ transition with almost pure $x-$polarisation. We apply an electric field of $1\,$GHz to lift the degeneracy between the $\vert E_{x}\rangle$ and $\vert E_{y}\rangle$ states, and also to increase the detuning between $\vert 0\rangle \leftrightarrow \vert M_{5}\rangle$ and other transitions. The electric field has a negligible effect on the ground state triplet, leading to an amplitude mixing of the $\vert \text{+1}\rangle$ and $\vert-1\rangle$ levels on the order of $2\times 10^{-5}$. The closest strongly allowed transition to $\vert 0\rangle \leftrightarrow \vert M_{3}\rangle$ is the $\vert \text{+1}\rangle \leftrightarrow \vert M_{5}\rangle\equiv(0.98\vert A_{1}\rangle+0.17\vert A_{2}\rangle)$ transition, with a detuning of $\delta_{\omega}=2\pi\times 2.71\,$GHz. Furthermore this transition is almost purely circularly polarised. Assuming transform-limited linewidth, at low temperatures (2 K) the excited-state decay transitions have amplitude decay rates of $\gamma(M_3)=2\pi\,\times\,6\,$MHz and $\gamma(M_5)=2\pi\,\times\,11\,$MHz \cite{Robledo2011a} so that in both cases $\delta_{\omega}\gg \gamma$. All other significantly allowed transitions are detuned even further and can be neglected.
\subsection{Quantum non-demolition measurement of the electron spin state}

We now consider an NV$^-$ centre placed at the antinode of a cavity resonant with the $\vert 0\rangle \leftrightarrow \vert M_{3}\rangle$ transition. The natural entanglement we can generate between the electron spin and optical field allows us to perform a quantum non demolition (QND) measurement \cite{Imoto} of the electron-spin (and thus also its initialisation). We can use a single photon to probe the electron spin a number of times and from the measurement patterns (clicks or no clicks) determine with high probability the state of the electron spin, that is whether it is in the $|0\rangle$  or $|+1\rangle$ state. In the following, we assume the cavity to have no losses other than the transmission through the mirrors, and a spatially perfectly mode-matched input beam. The core of the proposal is based on the different effects of the NV$^-$ centre being in the ground state $\vert 0\rangle$ rather than in state $\vert \text{+1}\rangle$ on light impinging on the cavity.  The resonator is assumed to have a finesse $\mathcal{F}$ and a $1/e^2$ mode intensity radius $w_C$, leading to a cooperativity of
\begin{figure}[htb]
\includegraphics[width=\columnwidth]{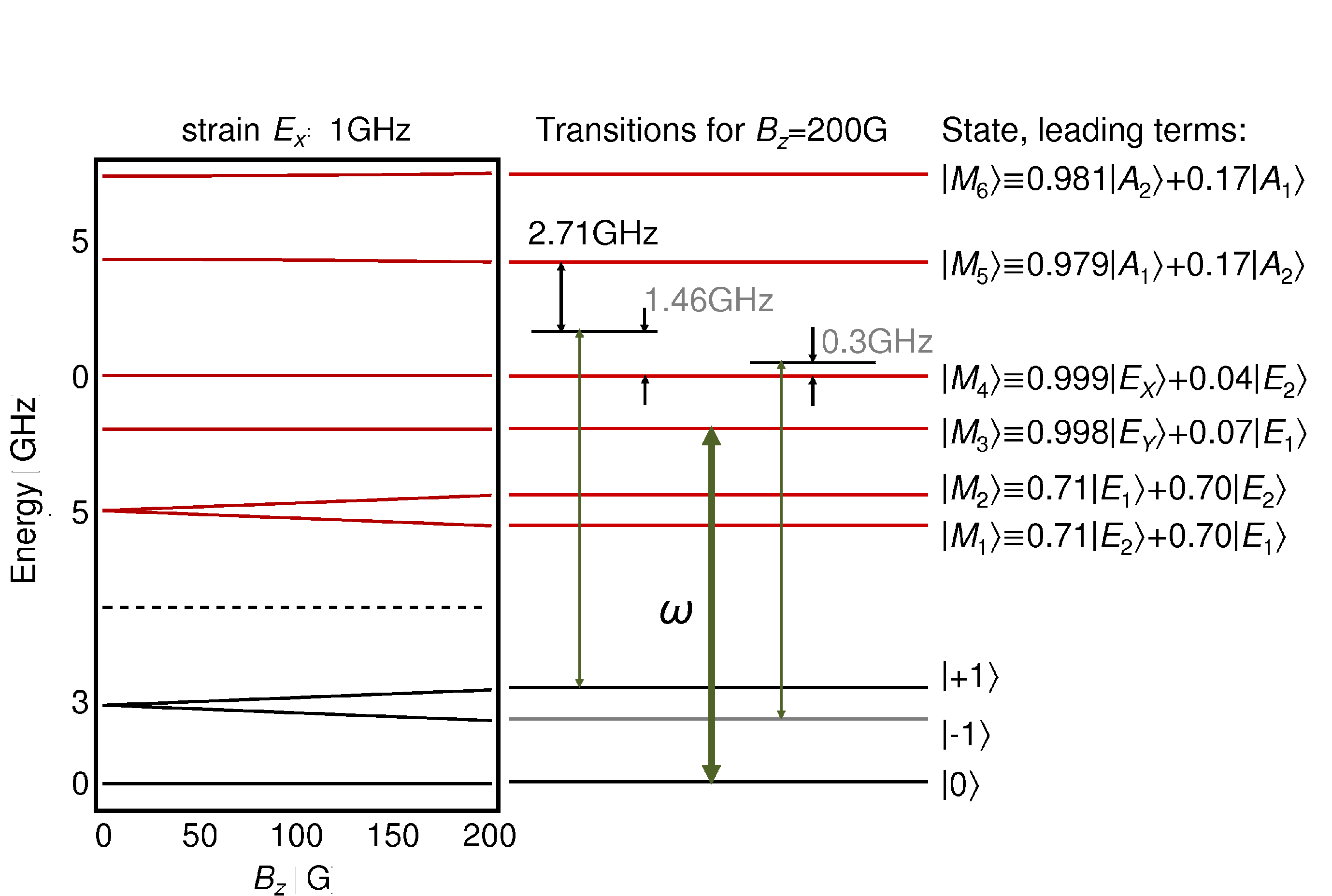}
\caption{\label{fig:states} Ground (black) and excited state energy eigenstates. Left: effect of $B_z$ up to the chosen field of $200\,$G, at an electric field $E_x=1\,$GHz. Middle: Level structure at the chosen fields, showing the laser frequency $\omega$ and its detuning from selected transitions. Right: leading terms of the energy eigenstates.}
\end{figure}
\begin{equation}\label{eq:coop}
C=\frac{2}{\pi}\frac{\sigma_E}{\sigma_C} \mathcal{F}\eta_{BR}\,.
\end{equation}
Here, $\sigma_E=3\lambda^2/(2\pi)$ is the emitter scattering cross-section, while $\sigma_{C}=\pi w_C^2$ is the cavity mode area. $\eta_{BR}$ is the branching ratio of the transition in question, which is $\eta_{BR}(M_3\leftrightarrow 0)=4\%$ and $\eta_{BR}(M_5 \leftrightarrow \pm 1)=2\%$.
The resonator amplitude decay rate depends on the resonator length $L$ with $\kappa=\pi c/(2L \mathcal{F})$. The reflection and transmission amplitudes for a photon with a linewidth $\gamma_{phot}\ll\kappa,\gamma$ being reflected or transmitted by a cavity containing an NV$^-$ centre are given by
\begin{eqnarray}
\label{eq:amps}
A_r &=& 1-\frac{1-A}{(1-i\Delta_C)-2C/(1-i\Delta_E)},\nonumber\\ 
A_t &=& \frac{\sqrt{1-A^2}}{(1-i\Delta_C)-2C/(1-i\Delta_E)},
\end{eqnarray}
with $\Delta_C=(\omega_{laser}-\omega_{\text{cavity}})/\kappa$, $\Delta_E=(\omega_{\text{laser}}-\omega_{\text{NV}})/\gamma$, and where $A=(r_1-r_2)/(1-r_1 r_2)$ is the amplitude of the reflected light for an empty cavity on resonance, with amplitude reflectance coefficients $r_{1,2}$ for the input and output mirrors, respectively. Then the probabilities for reflection and transmission are
\begin{eqnarray}
\label{eq:probs}
P_R &=& 1-\frac{(1-A)\left(1+A+4C+(1+A)\Delta_E^2\right)}{4C^2+4C(1-\Delta_E\Delta_C)+(1+\Delta_E^2)(1+\Delta_C^2)},\nonumber\\ P_T &=& \frac{(1-A^2)\left( 1+\Delta_E^2\right)}{4C^2+4C(1-\Delta_E\Delta_C)+(1+\Delta_E^2)(1+\Delta_C^2)}.
\end{eqnarray}
By energy conservation, the incoherent scattering probability is $P_S\left(\vert 0\rangle\right)=1-(P_R+P_T)$.
For emitter, cavity and probe light tuned to resonance, the expressions reduce to
\begin{equation}\label{eq:probsRes}
P_R^{res.}=\left( \frac{2C+A}{2C+1}\right) ^2,\, P_T^{res.}=\frac{1-A^2}{(2C+1)^2}.
\end{equation}

We now need to maximise the difference in reflected signal caused by an NV$^-$ centre in the ground $\vert 0\rangle$ state, which can be done in two ways:\\
 \begin{itemize}
\item  \textbf{High-cooperativity implementation:}\;\; In this approach, we minimise $A$ so that $A\simeq0$ and maximise $C$. Then the signal for the empty cavity results in  $P_R\simeq0$ while the signal for a cavity containing an NV$^-$ centre in the ground $\vert 0\rangle$ state tends to $P_R\simeq1$ for $C\gg 1$.  In this limit, the emitter excitation decreases with cooperativity as $P_S\rightarrow1/C$ \cite{Reichel2012}. However, there is a small off-resonant excitation of NV$^-$ centres in the $\vert \text{+1}\rangle$ ground state  which remains even for large cooperativity, and limits the performance of the device. 

Excitation of the NV$^-$ centre can be significantly decreased for either the $\vert 0\rangle$ or $\vert +1\rangle$ ground states by using an appropriately polarised optical field. In principle, the excitation for one of these states can be entirely turned off. In our situation we select a polarised field  to suppress excitation in the $|+1\rangle \leftrightarrow |M_5\rangle$ such that $P_S(|+1\rangle) \rightarrow 0$. 

This approach also requires careful matching of mirror reflectivities.\\

\item \textbf{Low-cooperativity implementation:}\;\; In this approach, we select a large negative value for $A$ ($A\simeq -1$) and tune $C$ such that $2C+A=0$. This can be arranged by choosing $r_2\gg r_1$. This approach is both more flexible and more readily achievable as we only require an initial cooperativity of $C(\vert 0\rangle \leftrightarrow \vert M_{3}\rangle)\geq0.5$. The cooperativity can then be reduced to $0.5$ by rotating the polarisation of the incoming photons away from the $x-$direction. Alternatively, the interaction strength between light and emitter can be decreased by detuning.  Assuming a value of $A=-1$, a detuning of $\Delta_E=\Delta_C=\sqrt{2C-1}$ leads to vanishing reflection probability. Conversely, the reflection probability approaches $A^2$ when the NV$^-$ centre is in the state $|+1\rangle$, as can be seen from Eqn.~(\ref{eq:probs}), so detecting a photon projects the NV$^-$ centre onto this state.  However, as this implementation is based on the conditional absorption of a photon, the performance is limited by spin-flip-inducing transitions. Experimentally, these have been observed to be on the order of $1\%$, which excludes this implementation for our purposes unless this issue can be addressed. It will nonetheless be suitable for initial demonstrations of the entangling mechanism. For our work we therefore focus on the high-cooperativity implementation.

\end{itemize}

\begin{figure}[htb]
\includegraphics[width=5cm]{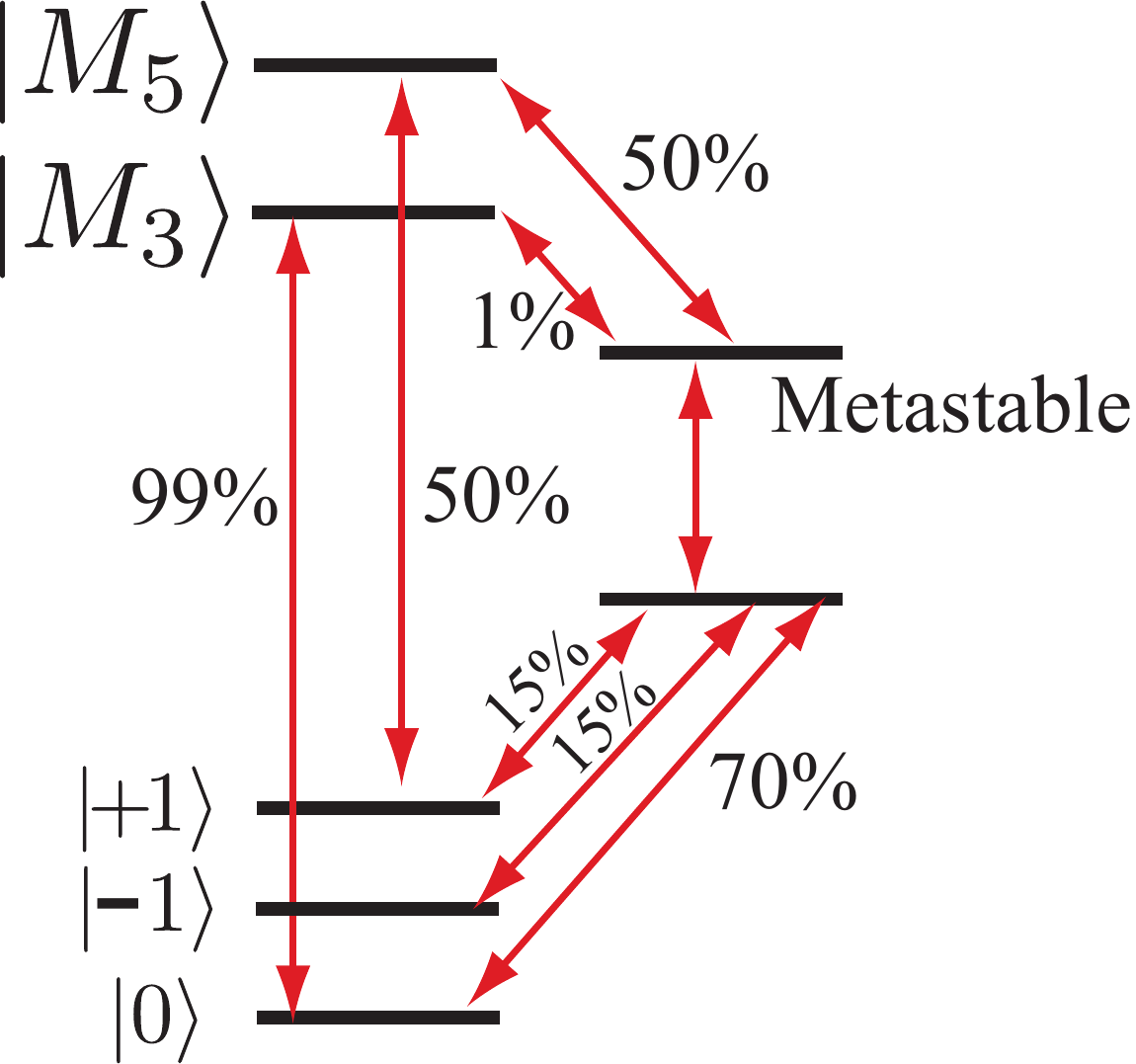}
\caption{\label{fig:mixing} Simplified energy level diagram with  key transitions through the metastable states. Indicated are approximate probabilities for these transitions occurring}
\end{figure}

{\bf Measurement sequence and sources of error}:  We aim for near perfect contrast of the empty cavity and maximum reflectivity---that is, $A_r(|+1\rangle)\sim A\rightarrow 0$ and $A_r(|0\rangle)\rightarrow 1$. Our state detection is based on a measurement -- spin rotation -- measurement sequence where we assume that negligible errors occur during in the spin rotation.  Furthermore, we assume that this sequence will be repeated many times, as photon loss will be unavoidable in a realistic device. If the electronic spin is in the $|0\rangle$ state, the probability that our single-photon detector clicks at least once in $s$ attempts is
\begin{eqnarray}
P_{click, 0}(s)&=& 1-(1-|\eta A_r(|0\rangle)|^2))^s \nonumber \\
&=& 1 - \sum_{i=0}^s     \binom{s}{i} (-1)^i |\eta A_r(|0\rangle)|^{ 2 i}.
\end{eqnarray}
Then, the probability of at least one click in $s$ attempts with no spin flips is 
\begin{eqnarray}
 \sum_{i=1}^{s }  |\eta A_r(|0\rangle)|^{2 i}  \left[1-|\eta A_r(|0\rangle)|^2-P_S(|0\rangle) P_{\text{flip}, 0} )\right]^{s-i} \nonumber,
 \end{eqnarray}
 where $P_{\text{flip}, 0}$ and $P_{\text{flip}, +1}$ are the probabilities of a single measurement inducing a spin flip upon excitation when the NV$^-$ centre is in one of the qubit states, due to resonant and off-resonant excitation, respectively.   Conversely, the probability for the detector never to click in $s$ attempts and not spin flip when the NV$^-$ centre is in the $|+1\rangle$ state, is 
 $$P_{click, +1}(s)=   \left[1-|\eta A_r(|+1\rangle)|^2- P_S(|+1\rangle)  P_{\text{flip}, +1}\right]^s$$ 
 where $\eta^2$ is the single photon detection efficiency, including all losses along the channel. The error probability for our entire sequence is then
\begin{align}\label{eq:QNDErr}
P_{err}^{QND}\sim &\; 1 - \sum_{i=1}^{s }  |\eta A_r(|0\rangle)|^{2 i} \nonumber \\
\times&	 \left[1-|\eta A_r(|0\rangle)|^2-P_S(|0\rangle) P_{\text{flip}, 0} )\right]^{s-i} \\
\times&    \left[1-|\eta A_r(|+1\rangle)|^2- P_S(|+1\rangle)  P_{\text{flip}, +1}\right]^s \nonumber .
\end{align}
One can immediately see the advantage of having $ |A_{r}(|+1\rangle)|^2 \sim A = 0$. A detector click strongly indicates that the NV$^-$ centre is in the $|0\rangle$ state. We assume $P_{\text{flip}, 0}=0.003$ and $P_{\text{flip}, +1}=0.35$ respectively (see Fig.~\ref{fig:mixing}), but we note that there is no current consensus on these values in the literature \cite{DohertyReview}. A key advantage of our measurement sequence is that it allows us to determine whether the NV$^-$ centre exits the qubit subspace into the state $|-1\rangle$.

The scheme can easily be modified to use weak coherent pulses instead of single photons.  We neglect this approach to avoid errors due to de-ionization of the NV$^-$ centre, which are possible for coherent states given their non-zero overlap with Fock states with $n>1$.  While the probability of this occurring may be small, it may be difficult to detect explicitly.

\begin{figure}
\includegraphics[width=\columnwidth]{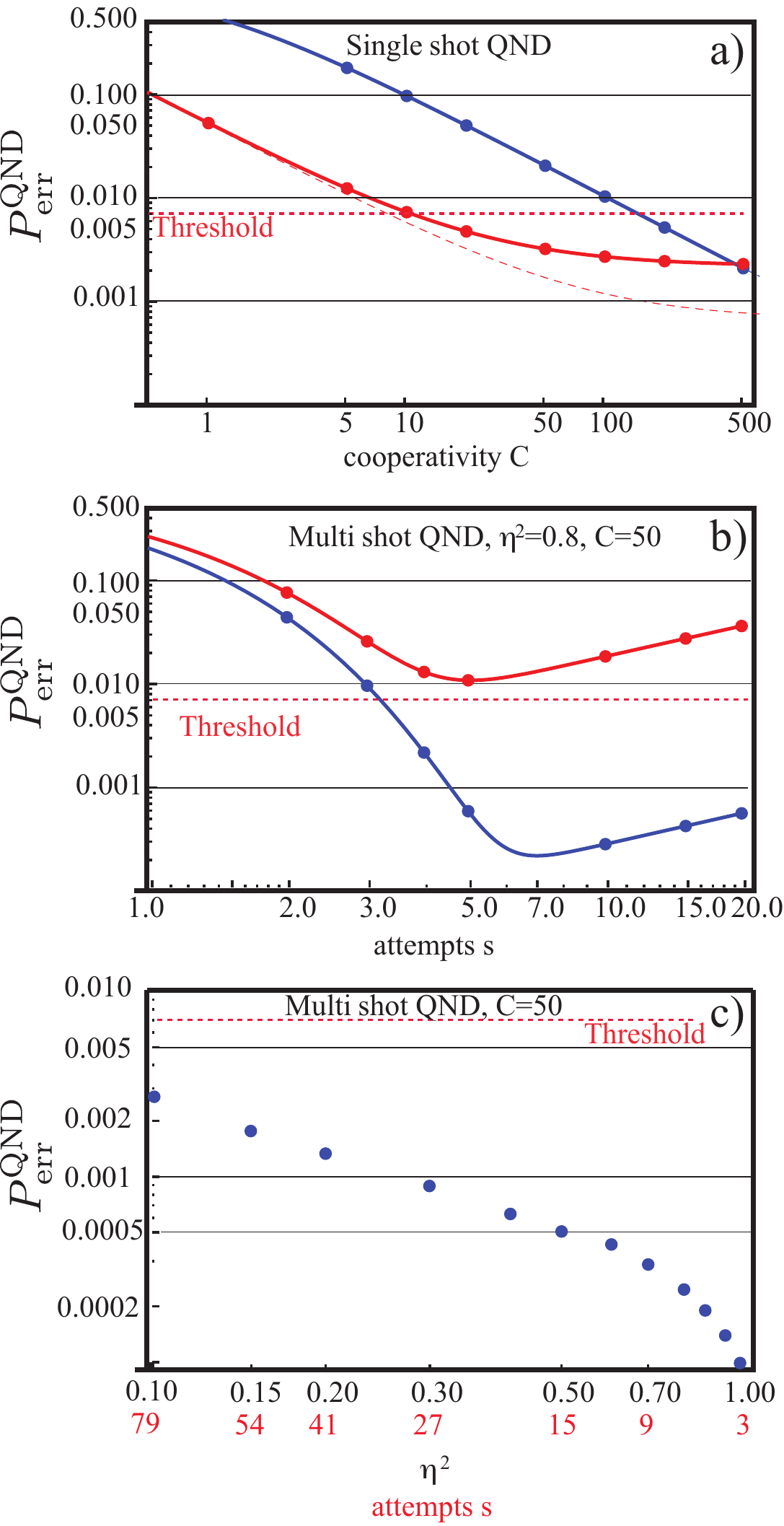}
\caption{\label{fig:qndError} a)  Single-shot failure probability for the QND measurement, including false detection results and spin flips during the measurement. Red: $A=-0.95$, blue: $A=0$. The plot assumes no detection losses or dark counts. Points are for numerically optimised detuning values, while solid lines are found by analytical approximation (see text). The dashed lines show the cases $P_{\text{flip}, 0}=0$, $P_{\text{flip}, +1}=0.35$ (red) and $P_{\text{flip}, 0}=0.003$, $P_{\text{flip}, +1}=0$ (blue).  b) Failure probability for multiple attempts $s$ assuming a detection efficiency of $\eta^2=0.8$ and a cooperativity of $C=50$. Dark counts are neglected. Here $P_{\text{flip}, 0}=0$, $P_{\text{flip}, +1}=0.35$. Points are for numerically optimised detuning values, while solid lines are found by analytical approximation (see text). c)   Numerically optimised error rate versus detection efficiency, showing the desired cooperativity and required number of attempts.}
\end{figure}

{\bf QND measurement performance}: The QND measurement performance is depicted in Fig.~\ref{fig:qndError} for both the  low-cooperativity and  high-cooperativity approaches.  The low-cooperativity approach ($A\approx -1$)  is limited by the spin-flip probability per measurement for the resonant excitation situation. The numerically optimised points for the case of negative $A$ are closely matched by choosing $\Delta_E=\Delta_C=\sqrt{2C-1}$ (red lines in Fig.~\ref{fig:qndError}), while for $A=0$, $\Delta_E=0$ and $\Delta_C=C\gamma(M_5)/\delta_{\omega}$ (blue lines in Fig.~\ref{fig:qndError}). The error rate of a single QND measurement is on the order of $1\%$. For our scheme, we require an error rate of less than $7.3 \times 10^{-3}$ (see main text). To overcome this limitation, we consider the high-cooperativity approach where we perform multiple measurements, shown in the middle panel of Fig.~\ref{fig:qndError}. The error rate as a function of detection efficiency is shown in the lower panel of Fig.~\ref{fig:qndError}. We require optical efficiency (including detectors) of $\eta^2\gtrsim 30\%$ to meet the requirements of our scheme (see main text). A useful working cooperativity is of the order of $C\sim 50$.

Before outlining how to generate remote entanglement, we will briefly discuss detection errors, namely photon loss and dark counts:
\begin{itemize}
\item {\bf Photon loss}: This is the most common error, which can arise from a number of sources including absorption or scattering in the channel, coupling inefficiencies between the cavity and channel, and inefficient single-photon detection. This error simply decreases the probability that we successfully measure a photon at the detector.  We can model this by a parameter $\eta^2$ which ranges from [0,1] with  $\eta^2=1$ being no loss. The probability of successfully measuring the photon is the ideal success probability multiplied by $\eta^2$.

\item {\bf Dark counts}: This error is where the detector clicks when no photon was incident. In principle, with current gated APDs, this dark count probability could be less than $10^{-5}$ per time window \cite{Dorenbos}. 
\end{itemize}
 
\subsection{Entanglement}
\begin{figure*}[htb]
\begin{center}
\includegraphics[scale=1.2]{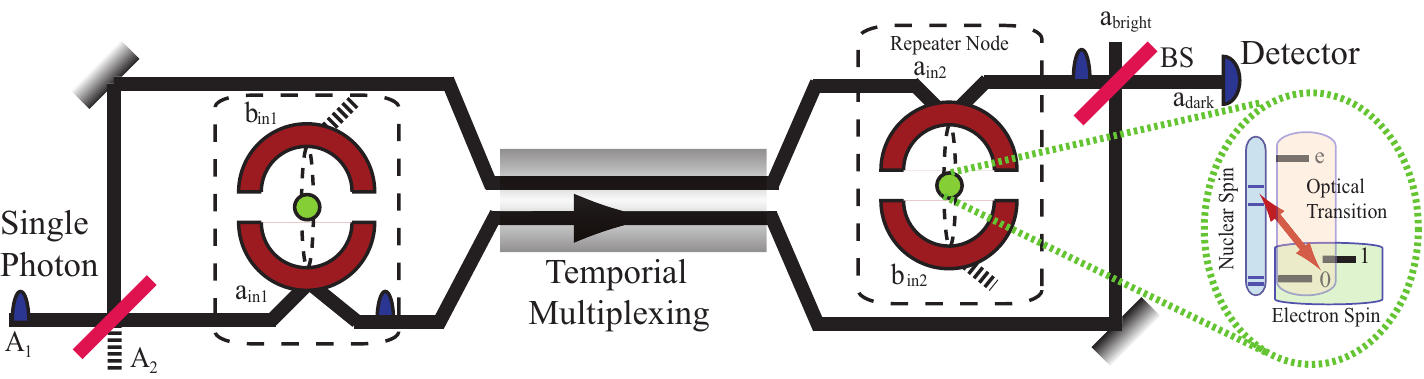}
\end{center}
\vspace{-10pt}
\caption{Schematic representation of the entanglement of two individual NV$^-$ centres located in remote cavities based on a single photon conditioning measurement. A single photon is split into two modes on a 50:50 beamsplitter with the bottom mode directed to the first cavity.  This mode, now containing a superposition of no photon and one photon, interacts with the electron spin prepared as $\frac{1}{\sqrt 2} \left[|0\rangle+|+1\rangle\right]$.  The change of transmission coefficient dependent on the electron spin state entangles our two subsystems. Then the reflected mode from the cavity and the top mode from the 50:50 beamsplitter  are temporally multiplexed into the same fiber and transmitted to the second module containing an NV$^-$ centre. The temporally multiplexed photonic signal is then separated back into its original two modes and the upper mode interacts with the NV$^-$ centre in the cavity.  The two modes are then recombined on a 50:50 beamsplitter and the dark port monitored. A photon detected at this port projects the two electron spins into the maximally entangled singlet state.} 
\label{spin-entangler} 
\end{figure*}

The creation of an entangled state between two remote electron states can be described in a straightforward manner given the previous discussion. Our scheme, which we depict in Fig.~\ref{spin-entangler}, is comparable to the protocol of Duan, Lukin, Cirac and Zoller \cite{dlcz}. We place two microcavities, each containing a single NV$^-$ centre, at the output ports of a 50:50 beamsplitter in a Michelson interferometer configuration. For simplicity, we set $A_{r,i}(|+1\rangle)=A_i$ and $A_{r,i}(|0\rangle)=A_{r,i}$ with $i=(a,b)$ the indices of the two cavities.  

In the most general case, we start our sequence by first preparing the NV$^-$ centres in superpositions $\alpha_a\vert \text{+1}\rangle _a+ \beta_a\vert 0\rangle _a $ and $\alpha_b\vert \text{+1}\rangle _b+ \beta_b\vert 0\rangle _b $. A single photon then impinges on the beamsplitter, resulting in a path-entangled state being sent to the two cavities.  The photon then interacts with the cavities and then returns to the beamsplitter. A detection event in the dark port projects the NV$^-$ centres into the state
\begin{eqnarray}
\psi_d&=&(A_{r,b} -A_{r,a})\alpha_a \alpha_b/\sqrt{2}\vert \text{+1},\text{+1}\rangle _{a,b} \nonumber \\
&\;& +(A_{b} -A_{a})\beta_a \beta_b/\sqrt{2}\vert 0,0\rangle _{a,b} \nonumber \\
&\;&+ (A_{b} -A_{r,a})\alpha_a \beta_b/\sqrt{2}\vert \text{+1},0\rangle _{a,b} \nonumber \\
&\;&+ (A_{r,b} -A_{a})\beta_a\alpha_b /\sqrt{2}\vert 0,\text{+1}\rangle _{a,b}. 
\end{eqnarray}
Now non-zero values of $A_{r_i}$ and $1-A_i$ will only lead to a decrease in the state amplitude, while differences between the two cavities will generally lead to a loss in fidelity. This can be seen from the first two terms in the state $\psi_d$. Assuming perfect state preparation with $\alpha_i =\beta_i =1/\sqrt{2}$, $A_a=A_b=0$ and $A_{r,a}=A_{r,b}=A_r$, our expression for $\psi_d$ simplifies to $A_r/\sqrt{8}(\vert 0,\text{+1}\rangle _{a,b}-\vert \text{+1},0\rangle _{a,b})$. The probability of projecting the two NV$^-$ centres onto our desired entangled state is $p_c=\eta^2 A_r^2/8$. This probability may seem quite low, however the Bell state is generated with extremely high fidelity, even with imperfect transmission and reflection coefficients. Instead of impacting the fidelity of the resulting singlet state, $A$ and $A_r$ impact the probability of detecting a single photon in the dark port.  

No photon detection leaves the electron in an indeterminate state, as the photons could have been lost in the channel, scattered from the NV$^-$ centres, or lost due to imperfect coupling, inefficient detection, or through the unmonitored $b_{{out}\;a,b}$ cavity ports. To address the low success probability, we repeat the process a number of times to establish a link with high probability \cite{barrett2005}.

So far we have assumed the transmission and reflection coefficients of the two cavities have been matched. This may not be the case in practice, and we will likely have $A_{b} \sim A_{a}$ as $A_{a}, A_{b} \sim0$ but $A_{r,b}  \neq A_{r,a}$. In this case, we can introduce a small loss element into the reflected path of the photon with the greater $A_{r,i}$ coefficient to effectively decrease its amplitude. Hence our resulting state is 
$\psi_d \propto  \beta_a \alpha_b \vert 0,+1\rangle _{a,b} - \alpha_a \beta_b\vert \text{+1},0\rangle _{a,b}$, as required. 
 
\subsection{A little determinism: adding a $^{15}$N nuclear spin}

With electron-spin initialisation and readout, the ability to generate remote entanglement, and a microwave driving field to perform electron-spin rotations, we essentially have all the operations required for distributed quantum computation and communication, particularly via the preparation and measurement of cluster states \cite{Raussendorf2001}. However, unsuccessful attempts to introduce additional qubits to the cluster state may destroy entanglement that has already been established, significantly increasing the resource overhead for low success probabilities \cite{kok2007}.  Adding a little determinism will decrease these requirements. 

An NV$^-$ centre in diamond possesses an electron spin and also a nuclear spin from the $^{15}$N atom. These couple naturally via the hyperfine interaction given by $H_{\rm e-n}$, which may allow us to add another qubit to the module. With a 20 mT field, the exchange interaction component is far off-resonance and so in an appropriate rotating frame we can write the effective interaction Hamiltonian as
\begin{eqnarray}
H_{\rm eff} =\hbar A_{\rm net} |+1\rangle \langle +1 | \otimes | \uparrow\rangle \langle \uparrow| ,
\label{cphase}
\end{eqnarray}
where $A_{\rm net}= A_{\parallel}+\frac{A_{\perp}^2}{2 \Lambda}$ with $\Lambda=D+ g_e \mu_B B-g_n \mu_n B$ being the detuning between the electron and nuclear spin levels. This interaction gives a fast and natural controlled-phase (CPHASE) gate, where the time to create a maximally entangled state is $t_{\rm max}= \pi/  A_{\rm net} \sim 165$ ns.  

To transfer quantum information between these systems, we require single-qubit operations on both the electron and nuclear spins. The electron spin rotations can be achieved using a  $\sigma_+$ polarised microwave driving field of the form $H^{rf}_{\rm Driving}=  \hbar \Omega_0 \left[ e^{i \phi} | +1\rangle \langle 0 | +e^{-i \phi} | 0\rangle \langle +1|\right]$ (in our rotating frame). With $\phi=\pi/2$, a $-\pi/4$ Y-rotation transforms $|0\rangle \rightarrow \frac{1}{\sqrt 2} \left[ |0\rangle+|+1\rangle\right] $ (a Hadamard-like operation)  in approximately 2 ns \cite{everitt2012,footnote1}. The nuclear spin rotation operation could similarly be achieved through driving the exchange part of the hyperfine coupling in 1 $\mu s$ \cite{everitt2012}. We hence have the operations required to construct gates that transfer the state of the electron spin to the nuclear spin and vice versa. These gates may also be used to initialise and measure the nuclear spin, via a projective measurement of the coupled electron--nuclear system. We will discuss the error channels in the electron--nuclear spin system once we have integrated all the elements.  

\section{A hybrid interface}

Next, we combine the basic operations between the optical, electron-spin, and nuclear-spin components in a protocol for generating entanglement between two remote nuclear spins. Care is required to ensure that the operations work as intended. For instance, coupling between the electron and nuclear spin is always on, meaning that failed attempts at electron--electron coupling could cause errors on the nuclear spins. 

We begin by preparing the electron (nuclear) spin in the $|0\rangle$ ($|n_+\rangle=\frac{1}{\sqrt{2}}(|\downarrow\rangle+|\uparrow\rangle)$) state. An accurate (sub-nanosecond) clock is started in the first module and the electron spin rotated to $|+\rangle=\frac{1}{\sqrt{2}}(|0\rangle+|1\rangle)$.  Two independent operations occur at this time:
\begin{itemize}
\item First, as soon as the electron spin is rotated from $|0\rangle$ to $|+\rangle$, the hyperfine interaction begins coupling the electron spin with the nuclear spin, according to $|+\rangle|n_+\rangle \rightarrow  \frac{1}{\sqrt{2}} |0\rangle|n_+\rangle+\frac{1}{2}|1\rangle \left[ |\downarrow\rangle+e^{iA_{\rm net} t} |\uparrow\rangle\right]= |\Psi(t)\rangle$. The resulting entanglement is periodic and oscillates between separable and maximally entangled with period $2\pi/ A_{\rm net}\sim 330$ ns. The oscillation stops when the electron spin is returned to a polarised state. Alternatively we can use a spin-echo technique to disentangle the electron and nuclear spins at any time. We know that after a time $t$ the state $|+\rangle|n_+\rangle$ has evolved to $|\Psi(t)\rangle$. Performing a spin-echo pulse and waiting a further time $t$ evolves our combined state to $\frac{1}{\sqrt{2}}  |+\rangle \left[ |\downarrow\rangle+e^{iA_{\rm net} t} |\uparrow\rangle\right]$. The electron and nuclear spins are disentangled with the electron spin returning to the original state and the nuclear spin evolving to $\frac{1}{\sqrt{2}} \left[ |\downarrow\rangle+e^{iA_{\rm net} t} |\uparrow\rangle\right]$.

\item Second, a single photon is split on a 50:50 beamsplitter into two modes. The bottom mode in Fig.~\ref{spin-entangler} interacts via dipole-induced transparency with the NV$^-$ centre in the first module, where it becomes entangled with the electron spin state. Both modes exciting the cavity are temporally multiplexed and transmitted over a fiber to the second module where the multiplexing is reversed. In the second module, the clock is started and the electron spin in the cavity is rotated to $|+\rangle$ where it interacts with the top mode from the original beamsplitter. The two modes are recombined on the beamsplitter and the dark port of the interferometer is monitored. 
\end{itemize}
Two possible outcomes, which we refer to as unsuccessful and successful, are distinguished by the measurement result:
\begin{itemize}
\item The unsuccessful case is where no photon is detected at the dark port, which occurs if a photon is detected at the bright port or not at all (it may have been lost in the cavity, during coupling, or in the channel, or the detector may not have detected it due to error). In this case, we are unsure of the exact state of the remote electron spins and must assume it is maximally mixed.  Consequently (assuming that $A_{\rm net}$ is identical for both NV$^-$ centres), the density matrix of the combined nuclear--electron system is 
\begin{eqnarray}\label{eq:density}
\rho &=& |00\rangle\langle00|_e \rho_n \nonumber \\
&+& |01\rangle\langle01|_e e^{-iA_{net}tZ_{n_2}}\rho_ne^{iA_{net}tZ_{n_2}}  \nonumber \\
&+& |10\rangle\langle10|_e e^{-iA_{net}tZ_{n_1}}\rho_ne^{iA_{net}tZ_{n_1}}  \nonumber \\
&+&  |11\rangle\langle11|_e e^{-iA_{net}tZ_{n_1}Z_{n_2}}\rho_ne^{iA_{net}tZ_{n_1}Z_{n_2}},
\end{eqnarray}
where $e$ and $n$ denote the electron and nuclear subsystems respectively.   The hyperfine coupling combined with the fact that photon loss completely mixes the state of the electrons implies that either one or two phase errors can be back-propagated to each nucleus. However, the nuclear component of this mixed state "re-purifies" itself with the periodicity $A_{\rm net}t = 2\pi m$ of the hyperfine coupling or via a spin-echo pulse (the spin-echo pulse is preferred as it potentially much faster). After such a pulse the electron and nucleus become decoupled and the state of the  nuclear qubits is simply $\frac{1}{\sqrt{2}} \left[ |\downarrow\rangle+e^{iA_{\rm net} t} |\uparrow\rangle\right]$. This slight phase rotation  $e^{iA_{\rm net} t}$, where $t$ is when the spin-echo pulse is applied, can be corrected later.

\item The successful case is where a photon is detected at the dark port and the remote electron spins are projected into a singlet state with a high fidelity, as discussed in Section IIB. A spin-echo pulse is also performed on each module to decouple the electron and nuclear spins. 
\end{itemize}

At this point, the electron and nuclear spins are decoupled. What to do next depends on the measurement result:
\begin{itemize}
\item In the unsuccessful case, we measure the electron spin at the $2t$ time of the spin echo and initialise the electron spin into $|0\rangle$, which collapses the overall density matrix to one of the four terms in Eqn.~(\ref{eq:density}). Then we start the procedure again from where we clocked the first attempt. Although the electron spin states are completely mixed, this gate sequence allows the nuclear spins to avoid decoherence and be preserved for the next attempt. This can be repeated until success. Errors may propagate to the nuclear spins due to poor control of the time when electrons are reinitialised to the $\ket{+}$ state prior to each attempt.

\item In the successful case, we perform a single-qubit  $\pi/4$ $Y$-rotation on one of the two electron spins at the $2 t$ time of the spin echo (the $\pi/4$ $Y$-rotation is an effective Hadamard gate necessary to convert the electron--electron singlet state  into the appropriate two-qubit cluster state, $(|0+\rangle - |1-\rangle)/\sqrt{2}$). We then wait until the hyperfine interaction maximally entangles the electron and nuclear spins within the node (at a time $t=m \pi/A_{\rm net}$). A second $\frac{\pi}{4}$, $Y$-rotation is performed on the electron spin of each module followed by measurement in the computational basis (an effective $X$-basis measurement).  
\end{itemize}

Upon success, we have transferred newly established entanglement between the electron spins in two remote modules to the nuclear spins in those same modules (by effectively teleporting a CPHASE gate), which is where we are storing and processing our quantum information. Importantly, the protocol circumvents photon-loss induced decoherence via the hyperfine interaction on the nuclear spin.

\subsection{Timescales}

The timescales for the various processes in the protocol can be grouped into three categories: short (1--30 ns), medium (100 ns -- 1 $\mu s$) and long ($>1$ $\mu s$). Short timescales are associated with electron spin operations (initialisation, detection, and rotations), medium timescales are associated with hyperfine coupling operations (entanglement, nuclear spin initialisation, and measurement in the $Z$-basis), and long timescales are associated with nuclear spin rotations (via the hyperfine interaction \cite{everitt2012}).  Nuclear spin rotations are generally only required only for initialisation, and the number of nuclear rotations is independent of the number of attempts to create an electron--electron bond. Similarly, measurement of the nuclear spin is only required for measurements that consume the preprepared entanglement. Transmission of a single photon between the modules is our last operation of interest, and its timescale depends on the task at hand. In quantum communication, remote modules may be separated by up to 40 km. In this case, it takes approximately $0.4$ ms to transmit a photon between modules and receive a classical return signal. In this case, the duration of each attempt is determined by this timescale. By contrast, for modules separated by 1 m, the transmission time is $\sim 10$ ns, which is shorter than the timescale associated with hyperfine coupling operations. 

The overall rate of the protocol is determined by the product of the per-attempt rate and the number of attempts. The number of attempts is related to the probability of success of each attempt, which depends on the efficiency of the optical components. We define the total efficiency of the optical components, $p_o$, to be the combined efficiency of all factors that influence the success probability of the optical gate, besides the theoretical upper bound of $0.125$. If $p_o=0.5$, then for each attempt the probability of success is $0.125\times0.5=0.0625$. After approximately 107 attempts the probability of success is $P=0.999$, which for our purpose is effectively deterministic.

\subsection{NV$^-$ module}
Let us now return to the issue of errors in the module. Error can be divided into two categories:
\begin{itemize}
\item Accumulation errors are those that depend on the number of attempts taken to establish entanglement between remote electron spins. These errors only affect the error rate of the nuclear--nuclear CPHASE gate, not the error rate of nuclear measurement and initialisation. To tolerate a low success probability (which necessitates a large number of attempts) these errors need to be heavily suppressed.
\item Non-accumulation errors are those that are independent of the total number of attempts, and depend only on the final successful attempt to establish entanglement between remote electron spins.
\end{itemize}
In Sections III and IV we will break down errors into several parameters that determine the overall error rates of nuclear spin measurement and initialisation CPHASE gates between remote nuclear spins.  These error rates will then allow us to determine the performance of the architecture.

\section{Nuclear Spin measurement and initialisation}
\begin{figure}[ht!]
\begin{center}
\includegraphics[scale=0.25]{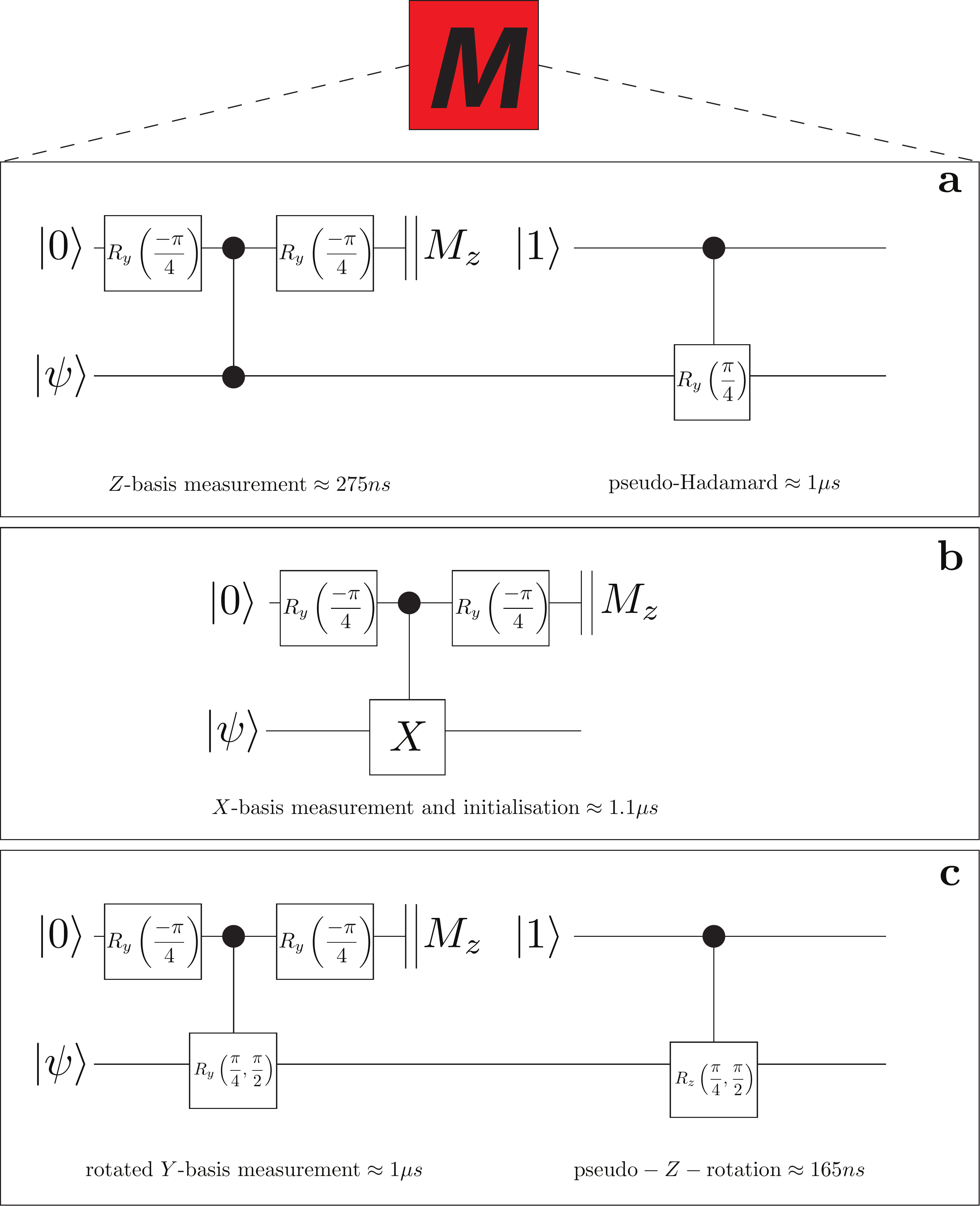}
\end{center}
\vspace{-10pt}
\caption{Projective measurement of the nuclear spin, mediated by the electron--nuclear hyperfine interaction. The natural hyperfine interaction enables fast $Z$-basis measurements, while a driven hyperfine interaction enables $X$ and $Y$ -basis measurements \cite{everitt2012}. (a) Measurement in the $Z$ basis and initialisation in $\ket{+}$ consists of measurement via the natural hyperfine interaction and then a controlled-$Y$ gate on the nuclear spin with the electron spin polarised in the $\ket{1}$ state, which effectively rotates the nuclear spin into the $\ket{+}$ state. 
(b) Measurement and initialisation in the $X$ basis.
(c) The two types of $Y$-basis measurements required by our scheme are performed by driving the hyperfine interaction. After measurement, a controlled $Z$-rotation with a polarised electron spin will reinitialise the nuclear spin in the $\ket{+}$ state.} 
\label{fig:meas} 
\end{figure}

Both measurement and initialisation of the nuclear qubit is performed via projective collapse of the electron and the hyperfine interaction.  As described in \cite{everitt2012} we can generate multiple types of controlled operations (where the electron acts as the control) between  the electron--nuclear system.  The most basic is the natural hyperfine generated CPHASE gate.  Combining this with $Y$  rotations on the electron, we are able to perform an effective $Z$-basis measurement on the nucleus with a total  time of approximately $2\times 5+165+100 = 275$ ns, assuming single-qubit gates take less than 5 ns and single attempt initialisation and measurement of the electron takes 100 ns (see Fig.~\ref{fig:meas}a).  This measurement circuit also initialises the nucleus in a known state.  Therefore, measurement and initialisation in this model is achieved with a combined gate.  

Similarly, we can drive the hyperfine interaction to generate a controlled rotation around a different axis (rather than the $Z$ axis).  Two examples are a controlled-not (CNOT) operation and a controlled-$Y$ operation, which can be used to measure the nucleus in the $X$ basis and $Y$ basis respectively (see Figs.~\ref{fig:meas}b and \ref{fig:meas}c).  Driving of the hyperfine interaction necessitates a longer time for these controlled operations (approximately 1 $\mu$s \cite{everitt2012}) and these are therefore classified as long timescale operations. Errors associated with nuclear spin initialisation and measurement do not depend on the number of attempts to establish entanglement between remote electron spins and may occur only when nuclear spins are measured.  

Gate times for the combined measurement and initialisation operation are approximately 1 $\mu$s.  The error rate associated with measurement and initialisation in the $Z$ and $Y$ basis is higher than in the $X$ basis as a second rotation is required to reinitialise the nuclear spin in the $X$-basis (to prepare cluster states, qubits should be initialised in $\ket{\pm}$).  As the natural CPHASE gate of the hyperfine interaction is much faster than the driven CNOT (or controlled-$Y$) gate, the timescale associated with measurement and initialisation in the $Z$-basis is the same as in the $X$-basis. 

\subsection{Intrinsic decoherence in diamond}
For both the nucleus and electron, intrinsic decoherence can be induced through spin relaxation (thermalisation) and through dephasing.  For the nuclear spin we can model both processes as  a Markovian process where the errors induced are approximately given by,
\begin{eqnarray}
p^{(1)}_{n}(t) &\approx& \left(1-e^{-t/T_{1n}}\right) \nonumber \\
p^{(2)}_{n}(t) &\approx&\frac{1}{2} \left(1-e^{-t/T_{2n}}\right) \nonumber,
\end{eqnarray}
where $t$ is the length of time considered and $T_{in}$ are the decoherence times ($i=1$ for relaxation and $i=2$ for dephasing).  Dephasing (from $T_2^*$ processes) results in $Z$ errors while relaxation (thermalisation)
 results in $X$ and $Y$ errors. We assume here that spin-echo techniques are being used on the electron spin to effectively decouple the electron and nuclear spins. If this is not the case, the coherence times of the nuclear spin will be much shorter. 

For the electron spin, relaxation can be modelled again as a simple Markovian process $p^{(1)}_{e}(t) \approx 1-e^{-t/T_{1e}}$, which results in $X$ and $Y$ errors. Dephasing is non-Markovian but gives $Z$ errors with probability $p^{(2)}_{e}(t) \approx \left(1-e^{-t^2/2T_{2e}^2}\right)/2$. Generally the relaxation times are very long (seconds) compared to the electron-spin gate times (nanoseconds to microseconds) and can be neglected, leaving only $Z$ errors as our intrinsic error. We use approximate expression for $p_{n,e}(t)$  to simplify our estimates and to find an upper bound for our error probabilities.  The master equation used for each process will give slightly different expressions for $X$, $Y$, and $Z$ errors.

Control errors can be modelled by an error $\epsilon$, which is defined as the over or under rotation caused by imprecise control of the Hamiltonian of the electron spin. Over or under rotation simply produces a error of the same type as the rotation axis, whereas axis misalignment may cause an arbitrary error. We assume that this error affects the rotation angle and not the rotation axis. In either case, given a rotation error of $\epsilon$, the error induced is given by $\sin^2(\epsilon) \approx \epsilon^2$, for $\epsilon \ll 1$.

The total error for electronic rotations (the combination of decoherence and control errors) is given by
\begin{equation}
p _e(5ns)=\frac{1}{2} (1-e^{-5ns^2 /T^2_{2e}})+ \epsilon^2.
\end{equation}
A similar expression can be derived for $p_n$ for pure decoherence over time $t$,
\begin{equation}
p_n(t) = (1-e^{-t/T_{n}}).
\end{equation}
Note that we do not include an $\epsilon^2$ term for the nuclear spin error. This is because all nuclear rotations are achieved by driving the hyperfine interaction, where the associated error is given by the coupled error terms $p_{2z}$ and $p_{2x}$. These intrinsic errors associated with the hyperfine control may introduce correlated errors.  A more detailed analysis of these processes  can be found in \cite{everitt2012} and the total probability of error during these rotations will encapsulate both the systematic errors and the intrinsic decoherence on both the electron and nucleus over the relevant time scales.  These can be modelled by a general two-qubit depolarising map with probabilities $p_{2z}$,  $p_{2x}$ and $p_{2y}$. Each of these expressions can now be used to bound the error rate associated with nuclear measurement and initialisation,
\begin{equation}
\begin{aligned}
&p_{M_Z} = 2p_e+2p_M+p_{2x}+p_{2y}\\
&p_{M_{X}} = 2p_e + p_M +p_{2x}\\
&p_{M_{Y}} = 2p_e+2p_M+p_{2x}+p_{2y},
\end{aligned}
\end{equation}
where $p_e$ is the electronic rotation error, $p_M$ is electronic measurement error (also initialisation) and $p_{2(x,y,z)}$ are the errors associated with the hyperfine coupling for natural ($z$) or driven $(x,y)$ evolution. The timescale of each measurement is approximately 1 $\mu$s and the errors in $p_{M_Z}$ will be dominant.

\section{Electron--electron connection}

Errors that accumulate as we attempt to establish entanglement between remote electron spins arise from three sources:
\begin{enumerate}
\item {\it Hyperfine interaction timing errors}. After an attempt to entangle two remote electron spins, the hyperfine interaction must be allowed to evolve (including spin-echo sequences) to the $2\pi$ point so that the electron and nuclear spins are disentangled prior to the next attempt. If there is an associated timing error, $\nu$, a $Z$ error will propagate back to the nucleus with a probability of $\sin^2(\nu/165\;\textrm{ns}) \approx (\nu/165\;\textrm{ns})^2$. In Table \ref{tab:numbers} we give the required accuracy for a successful connection probability of $P=0.99$ (accumulated nuclear spin error of 1\%) and $P=0.999$ (accumulated nuclear spin error of 0.1\%) for various optical component efficiencies, $p_o$. The probability of the connection being successful using a single sided-cavity protocol is given by $p_c = 0.125p_o$.  

\begin{table}
\begin{center}
\vspace*{4pt}   
\begin{tabular}{|c|c|c|}
Optical efficiency & \multicolumn{2}{|c|}{Timing accuracy (connection attempts)} \\
$p_o$ & $P=0.99$ & $P=0.999$\\
\hline
100\% &2.81 ns (35) & 725 ps (52)\\
80\% & 2.5 ns (44) & 644 ps (66)\\
50\% & 1.95 ns (71) & 504 ps (107)\\
20\% & 1.22 ns (182) & 315 ps (273)\\
10\% & 0.86 ns (366) &  222 ps (549)\\
\end{tabular}
\caption{Timing error and number of attempts required to establish entanglement with probability (and fidelity) $P=0.99$ and $P=0.999$. Since the timing error accumulates with each attempt, there is a tradeoff between optical efficiency and timing accuracy.} 
\label{tab:numbers}
\end{center}
\end{table} 

\item {\it Nuclear decoherence.} As entanglement is established between remote electron spins over a series of attempts, decoherence will accumulate on the nuclear spins. Long nuclear decoherence times are required to accommodate the low success probability. We assume that the physical separation between NV$^-$ centres is short enough such that the optical protocol can be confirmed to have succeeded or failed within the 165 ns required for the electron--nuclear hyperfine gate. An unsuccessful attempt takes approximately $2\times 45+100+5 \sim 200$ ns (initialisation of the electron via measurement, rotation of the electron, and spin-echo to disentangle the electron and nuclear spins prior to 
the next attempt). Therefore, the nuclear decoherence will be $p_n(200\;\textrm{ns}) = (1-e^{-2.00\times 10^{-7}/T_{n}})$ per attempt. For $s$ attempts, this becomes $ p_n(s 200\;\textrm{ns})$.

\item {\it Excitation of the electronic system.} When attempting to entangle remote electron spins, or when measuring and initialising the electronic spin via an optical photon, we may accidentally excite the electronic system. When this occurs, the attempt is automatically unsuccessful as the photon has been absorbed. With high probability, the excited system will relax to its original state with no error induced on the nuclear spin. However, due to level mixing in the upper manifold, there is a possibility of a series of non-spin conserving transitions back to the ground state. As soon as the spin state of the electron changes, the timing control that we use to prevent errors back-propagating to the nucleus becomes unreliable. This error channel is active not only during every connection attempt, but when measuring and initialising the electron spin.

Experiments to precisely determine the relevant branching ratios for the decay of the electron have not been performed, but we can approximate these values using a theoretical model. Consider the basic level structure of the NV$^-$ centre shown in Fig.~\ref{fig:mixing}. The probability of a photon being absorbed by the NV$^-$ centre can be calculated using Eqn.~(\ref{eq:probs}). Given the parameters we assume for our system,
\begin{eqnarray}
&P_S(\ket{0}) = 0.0098, \; P_S(\ket{1}) \sim 0 \\
&P_R(\ket{0}) = 0.980, \; P_R(\ket{1}) = 1.7\times 10^{-7},
\end{eqnarray}
where $P_R$ is the probability of reflection for each state and $P_S$ is the probability of absorption for each state. The probability of error on the nuclear spin depends on the state of the electron spin, and the worst case is when the electron is in the $\ket{0}$ state (as the probability of excitation is higher). The probability of error on the nuclear spin also depends on the likelihood of an excitation causing a spin flip in the NV$^-$ centre.  The general error mapping, in the worst case, is given by
\begin{eqnarray}
\begin{aligned}
\rho &= \frac{P_0}{2}(|0\rangle\langle0|_e +|1\rangle\langle1|_e )\\
&+P_1\rho_n|0\rangle\langle0|_e \\
&+ P_2Z_n\rho_nZ_n|1\rangle\langle1|_e,
\end{aligned}
\end{eqnarray}
where $P_0$ is the probability that no absorption takes place and the photon is lost through other mechanisms.  $P_1$ is the probability that the NV$^-$ centre relaxes to the $\ket{0}$ state via a series of spin-0 levels and $P_2$ is the probability that it relaxes to the $\ket{+1}$ state when initially in the $\ket{0}$ state.  When the system relaxes to the $\ket{+1}$ state, the probability of a error on the nuclear spin is related to exactly when the electron decays from the meta-stable state back to the $\ket{+1}$ state with respect to the 165 ns $\pi$ point of the hyperfine coupling. Reliable estimates for this decay are not experimentally available, so we will attempt to make a large overestimate. If this decay pathway occurs, we assume that a full $Z$-error occurs on the nuclear spin. Each probability can be estimated from the probability of absorption, $P_S$, and the relative probabilities of each of the transitions,  
\begin{eqnarray}
\begin{aligned}
P_0 &= 0.9902.\\
P_1 &= P_S(\ket{0})\times (0.99+.01\times 0.7) = 0.0098\\
P_2 &= P_S(\ket{0})\times (0.01\times 0.3) = 2.9\times 10^{-5}.
\end{aligned}
\end{eqnarray} 
Therefore, $P_2$ is our estimate of the probability that an error occurs on the nuclear spin due to excitation of the electron spin. This is likely to be is a significant overestimate as we have not accounted for the timing of the electron relaxation relative to the $\pi$ point of the hyperfine interaction. This estimate was done assuming a cooperativity of $C=50$. By doubling this cooperativity, the probability of error halves.
\end{enumerate}

\section{Topological cluster states}

We will now outline how our protocol to establish entanglement between remote nuclear spins enables scalable quantum information processing. In particular, we will outline how to prepare cluster states appropriate for universal quantum computation and quantum communication. A common way to prepare cluster states involves two-qubit CPHASE gates between neighbouring qubits in some geometry \cite{kok2007}, and our protocol is effectively a CPHASE gate between remote nuclear spins. Therefore, with a cluster state stored in the states of the nuclear spins, our protocol can be applied with additional modules to introduce additional qubits to the cluster. In this way, we can prepare an arbitrary cluster state by repeating the protocol as required with a sufficient number of modules. 

\subsection{Topological cluster-state error correction} 

\begin{figure}[ht!]
\begin{center} 
\resizebox{80mm}{!}{\includegraphics{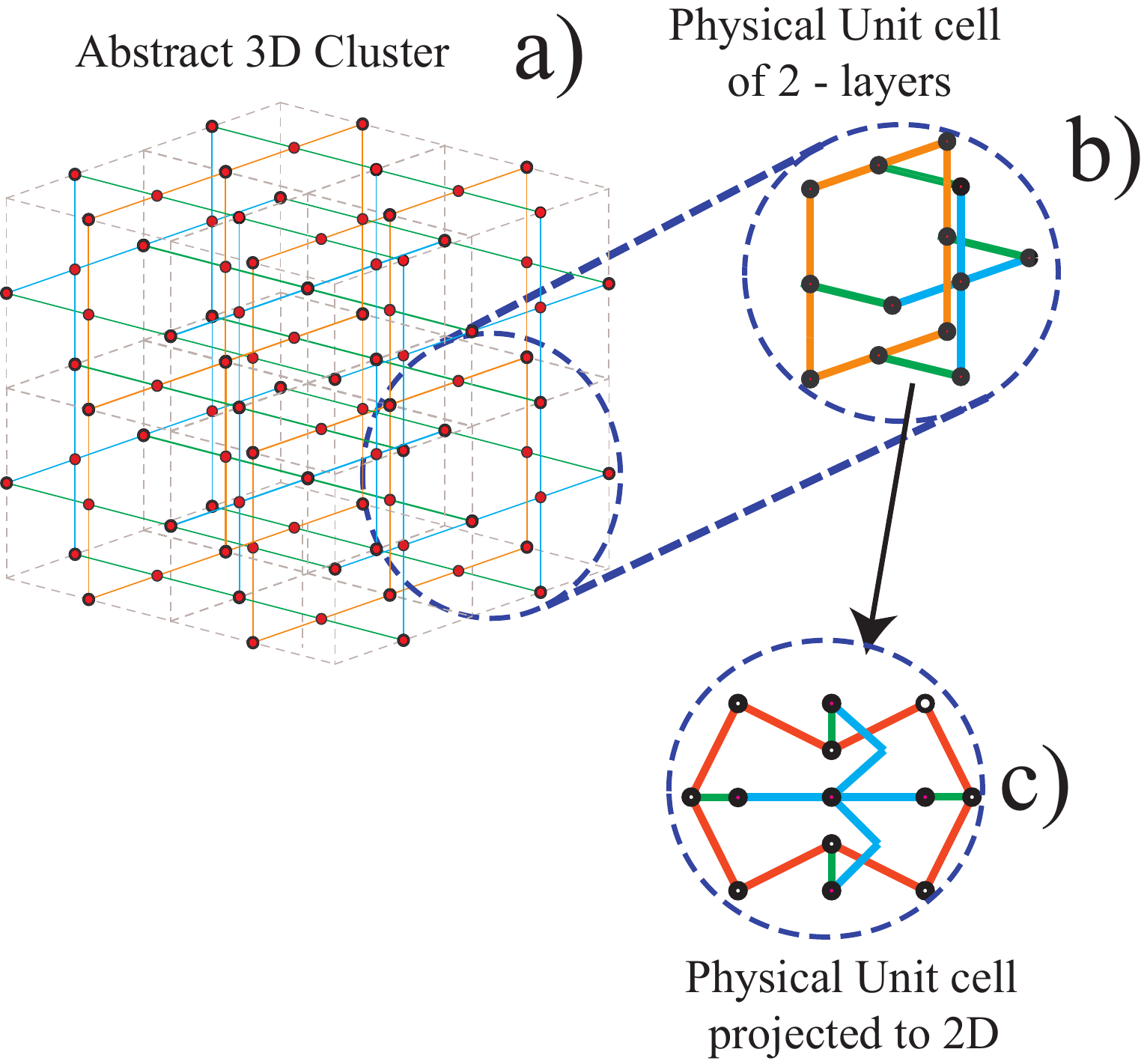}}
\end{center}
\vspace*{-10pt}
\caption{Schematic representation of a three-dimensional topological cluster state. a) A $2\times2\times2$ region of the topological cluster state. b) A physical unit cell comprising two layers of qubits. c) The projection of the physical unit cell to a two-dimensional plane, requiring nearest- and next-nearest neighbours interactions.}
\label{fig:cluster}
\end{figure}

For scalable quantum information processing, some form of error correction will be essential. Of the many schemes for error correction, the two-dimensional surface code and the closely related scheme based on three-dimensional topological cluster states are the strictly local schemes with the highest tolerance to errors (above $0.5\%$ per gate in both cases) \cite{RH07,RHG06,RHG07,WFSH10,BS10}. In both cases, each qubit is only required to interact with its four nearest neighbours. Typically, the surface code is thought to be appropriate for matter-based qubits, while topological cluster-state error correction is thought to be appropriate for photonic qubits. Despite the fact that our nuclear spin qubits are immobile, topological cluster-state error correction features a natural mechanism to tolerate missing bonds in the cluster state, which might arise in our scheme due to the strictly non-deterministic nature of the CPHASE gate. Missing bonds can be avoided through a clever interpretation of the measurement results during computation, at the cost of a reduced tolerance to other errors \cite{BS10}. This is not possible with surface code \cite{barrettstace10}. As such, we will focus on topological cluster-state error correction, which requires us to prepare the three-dimensional topological cluster state illustrated in Fig.~\ref{fig:cluster}a.  

In topological cluster-state error correction, two dimensions of the topological cluster state are reserved for the spatial distribution of protected logical qubits. The third dimension is identified with the temporal axis of the computation. As such, we are not required to prepare the entire topological cluster state before the computation can begin. Instead, only two adjacent layers of the topological cluster state are required at a given time. In Fig.~\ref{fig:cluster}b we illustrate the physical unit cell of the topological cluster state, comprising two layers. The back layer contains eight qubits connected in a square (orange), while the front layer contains five qubits connected in a cross (blue). The two layers are connected in the temporal direction (green). This pattern is repeated over the entire topological cluster state. Then, measurement of the front layer will teleport the current state of the computer to the back layer, at which point the front qubits can be reconnected in accordance with the geometry of the topological cluster state and the information can be teleported back again. In this way, the two physical layers  function as even and odd layers in the temporal direction, allowing an arbitrarily deep computation to be performed with a fixed number of physical qubits.

\subsection{Mapping to a two-dimensional geometry}

Because we are using matter qubits, it may be useful for the array of NV$^-$ modules to be strictly two-dimensional. In Fig.~\ref{fig:cluster}c we illustrate physical unit cell (comprising two layers in the temporal direction) projected to a two-dimensional plane (where colour coding has been preserved). Each NV$^-$ module is no longer connected to only its nearest neighbours, and several next-nearest neighbour connections are required. However, as these connections are optically mediated, this is compatible with our scheme. In principle, the array can be distributed, where neighbouring NV$^-$ modules are separated by an arbitrary distance (subject to photon loss and communication time) and the relevant integrated (or bulk) optics are positioned between connected modules.

\subsection{Connection circuits}

\begin{figure}[htb!]
\begin{center}
\resizebox{55mm}{!}{\includegraphics{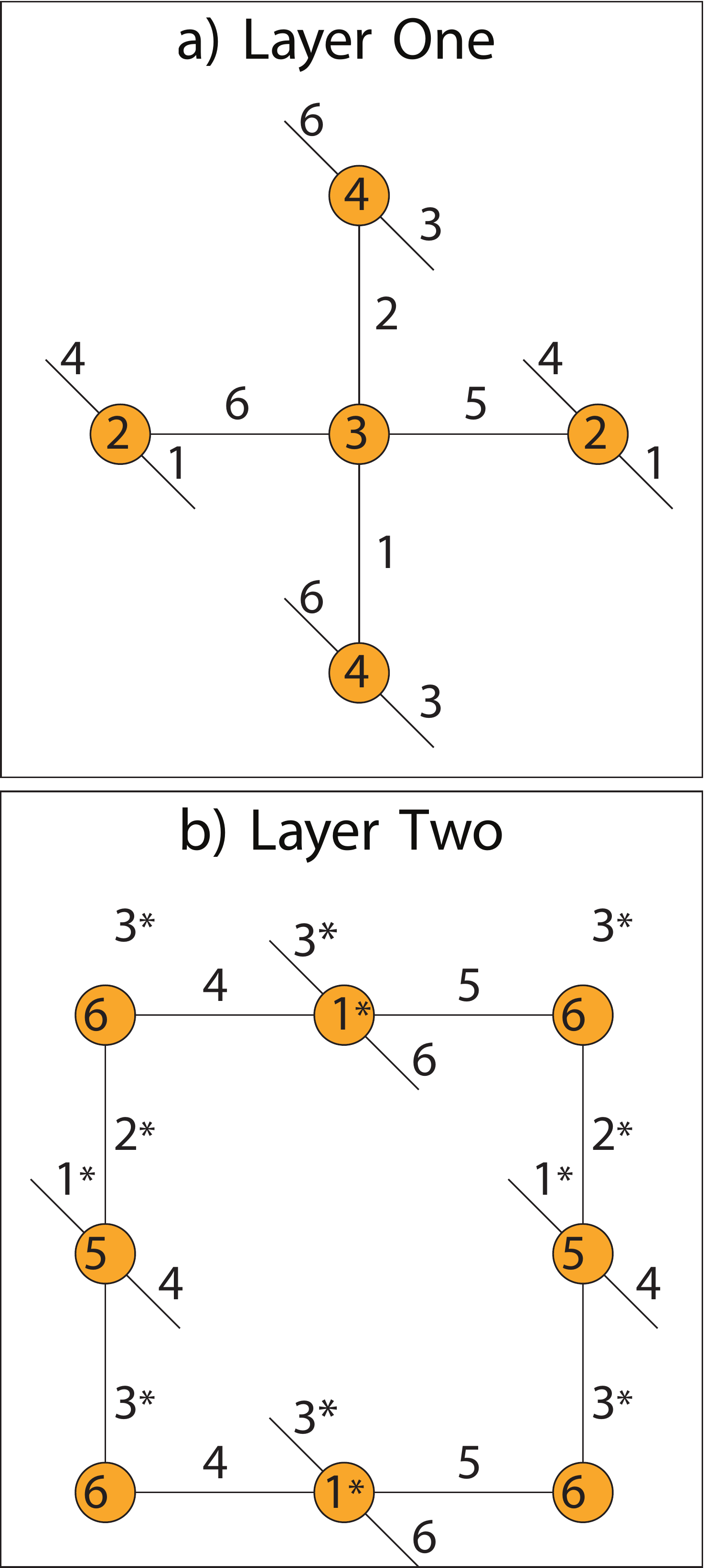}}
\end{center}
\vspace*{-10pt}
\caption{Sequence of NV$^-$ node connections for a unit cell of the physical cluster.  Each number represents the time-step for bonding, while a number inside each node represents measurement/initialisation.  This sequence is time optimal given the physical constraints on the system.  The star notation denotes equivalent time steps in the circuit which occur at later physical times, i.e. $1^*$ would occur at time step seven in real time and 
after the measurement a node remains idle of one step in the cluster.}
\label{fig:sequence}
\end{figure}

\begin{figure}[htb!]
\begin{center}
\resizebox{85mm}{!}{\includegraphics{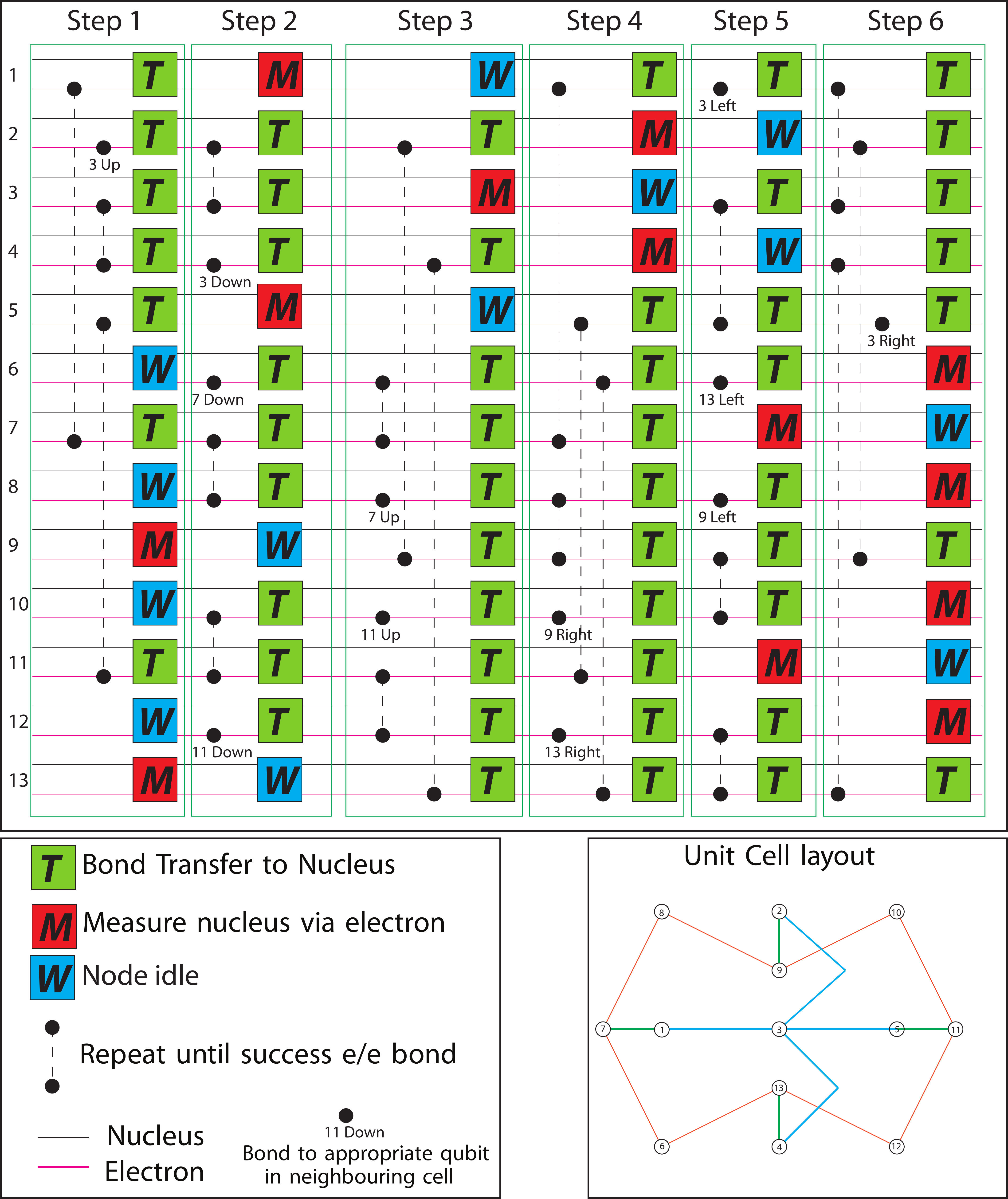}}
\end{center}
\vspace*{-10pt}
\caption{Quantum circuit required for the creation of two layers of the Raussendorf cluster.   Also shown are the circuits for nuclear initialisation and readout, utilising a QND measurement via the electron/nuclear hyperfine interaction.}
\label{fig:circuit}
\end{figure}

The circuit in Figs.~\ref{fig:sequence} and \ref{fig:circuit} is used to prepare the topological cluster state, layer by layer, with the array of NV$^-$ modules. Creating an optimal five-step circuit is not possible given only only two layers of modules. Instead, we use a six-step circuit, where NV$^-$ modules are idle for one step after measurement. In Fig.~\ref{fig:sequence}, the star notation denotes the subsequent six-step circuit that occurs at a later time (for example, $1^*$ denotes step $7$). Figure~\ref{fig:circuit} illustrates the circuit to prepare a topological cluster state with cross section equal to $1\times 1$ and arbitrary depth. As discussed in the main text, our calculation of the threshold assumes a five-step circuit. This is a reasonable approximation to the six-step circuit, as the error that accumulates while a module is idle is restricted to pure nuclear decoherence, which is negligible over the timescale of a successful electron--electron connection.


Simultaneous connections are grouped into a single step. In order to maintain synchronicity over the entire computer, all connections in a given step should be established before moving onto the next one.  As the connections are probabilistic, this may require some modules to wait while other modules are still being connected. However, as nuclear decoherence rates are orders of magnitude less than the time required to attempt an electron--electron connection, this waiting period will not adversely effect the error performance of the computer provided electronic errors propagating through the hyperfine interaction are handled carefully. This requirement determines the number of connection attempts, $g$, required for each module at a given success probability. The number of attempts to ensure a bond is established with probability $P$ is given by $g=\log(1-P)/\log(1-p_c)$, where $p_c$ is the probability that a given connection attempt is successful. Assuming that $p_c = 6.25\%$, $g=107$ for $P=0.99\%$. However, in the main text we assumed that $g$ is equal to the average number of connection attempts, given by $1/p_c$, which increases the rate of operation of the computer. In this case, we must ensure that the topological cluster state is synchronised over the entire computer.  On average, each node will synchronise with its neighbours. In extreme cases, some modules will have to wait for $g$ attempts to be connected. Our estimates will assume both the synchronous and asynchronous modes of operation. 

Because failed connections are heralded, we can exploit the tolerance of the topological cluster state to missing bonds \cite{BS10}. For example, we may reduce $P$ to 95\% to reduce the number of attempts. However, as the proportion of missing bonds is increased, the threshold error rate for all other errors is reduced. In our calculations, we do not exploit this potential robustness, and detailed calculations determining the tradeoff between missing bonds and other error rates will be studied in further work.
\section{Experimental requirements and expected performance}

We now estimate the experimental requirements for a scalable quantum computer based on our architecture. We have determined the threshold error rate for topological cluster-state error correction with a five-step circuit to be $\approx$ 0.73\%, as shown in the main text. This threshold is the maximum tolerable error rate for measurement/initialisation and CPHASE gates during preparation of the topological cluster state.

\begin{table*}
\begin{center}
\vspace*{4pt}   
\begin{tabular}{ccc}
\hline\hline
\hspace{0.2cm} Gate \hspace{0.2cm} & 
\hspace{0.2cm} Timescale \hspace{0.2cm} & 
\hspace{0.2cm} Error rate \hspace{0.2cm} \\
\hline
Single-qubit gate: electron & 5 ns & error, $p_e$ \\
 Initialisation/measurement via measurement: electron & 100 ns & incorrect state, $p_M$ \\
 Initialisation/measurement: Nuclear ($Z$) & $\approx 1.3$ $\mu$s & projective circuit, $p_{I,M_z}$ \\
 Initialisation/measurement: Nuclear ($X$) & $\approx 1.1$ $\mu$s & projective circuit, $p_{I,M_x}$\\
 Initialisation/measurement: Nuclear ($Y$) & $\approx 1.2$ $\mu$s & projective circuit, $p_{I,M_y}$\\
Electron--electron C-$\sigma_z$ error & $g(200\;\textrm{ns})$& $3p_{2z}+3p+p_e(275\;\textrm{ns})+p_M$  \\
Timing error: hyperfine interaction & & $(\nu/165\;\textrm{ns})^2$ \\
Nuclear--nuclear C-$\sigma_z$ & $g(200\;\textrm{ns})+110+165=(200g+275)$ ns & $p_{CZ}$\\
\hline\hline
\end{tabular}
\caption{Estimate of physical parameters and basic module operations.}
\label{tab:volumes}
\end{center}
\end{table*} 

To quantify the experimental requirements of the system, we specify several parameters in Table \ref{tab:volumes}, assuming that the time required for all connections is determined by the slowest connection, given by $g=\log(1-P)/\log(1-p_c)$. The average number of attempts for a successful connection is $1/p_c$. We assume that the accumulated error per connection is given by $E/p_c$, where $E$ is the accumulated error per attempt [excluding nuclear decoherence, which will induce an error of $gp_n(200\;\textrm{ns})$ as all of the nuclear spins are waiting until all connections have been made]. Then, the probability of error for measurement and initialisation of the nuclear spins and CPHASE gates between nuclear spins is
\begin{eqnarray}
p_{I,M_Z} &=& 2p_e+p_M+p_{2z}+p_{2y} + 2sP_{2}\\
p_{I,M_{X}} &=& 2p_e + p_M +p_{2x} + 2sP_{2}\\
p_{I,M_{Y}} &=& 2p_e+p_M+p_{2z}+p_{2y} + 4sP_{2}\\
p_{CZ} &=& \frac{((\nu/165\;\textrm{ns})^2+2sP_{2})}{p_c}+gp_n(200\;\textrm{ns})  \\
&+&3p_{2z}+3p_e+p_{e}(275\;\textrm{ns})+p_M+2sP_{2}. \nonumber
\end{eqnarray}
These expressions include errors associated with electronic rotations and hyperfine gates, and an additional term, $P_{2}$, which represents the probability that one of the $2s$ photons that are used in the QND measurement is absorbed by the NV$^-$ electron. 

For $p_{CZ}$, the first three terms correspond to the errors accumulated during $g$ connection attempts, while the remaining five terms are associated with the final successful connection.  We assume that for the vast majority of nodes, the number of attempts is $1/p_c$ and hence the 
accumulation errors from photon absorption ($P_2$) and the hyperfine interaction ($\nu$) are amplified by $1/p_c$.  However, the nuclear error 
is amplified by a factor of $g$ under the assumption that a given node will have to wait for the entire computer to synchronise.  The $p_e(275\;\textrm{ns})$ term is included because, after a $90$ ns spin-echo pulse, the successful connection will undergo a $Y$-rotation and be stored in the electrons for the $\pi = 165$ ns cycle time of the hyperfine interaction before the electron is measured and the connection is transferred to the nucleus (taking 105 ns). We have one set of errors for measurement/initialisation, the timescales for these gates are commensurate, and the majority of measurements in topological cluster-state error correction (where measurement errors are relevant) are $X$-basis measurements.  

These probabilities must satisfy the threshold condition of topological cluster-state error correction for the architecture to be scalable. For a useful device, they should be approximately an order of magnitude lower than the threshold, otherwise the resources required for error correction will become prohibitively large). As the threshold is estimated to be 0.73\%, we will require that $p_{I,M}$ and $p_{CZ}$ are both $\leq 0.1\%$.

Our expressions for $p_{I,M}$ and $p_{CZ}$ upper bound the total error rate for each operation. In the main text, we outlined the individual requirements for each physical parameter. A detailed simulation is required examine the full parameter space to find the optimal set of physical parameters such that each operation satisfies the threshold condition with the lowest possible error rate. In our simulations, we assume that $P_2 \neq 0$, hence the fidelity asymptotes to a value below unity as each individual error approaches zero. We can provide a set of parameters that can satisfy the threshold condition, but are not necessarily optimal. Assuming a success probability per attempt, $p_c$, of 6.25\%, the following set of error parameters meet the threshold condition:  
\begin{equation}
\begin{aligned}
\nu &= 0.05, \quad ( \text{50ps timing error}) \\
\epsilon &= 5\times 10^{-3}\\
T_{2e}^* &= 90\mu s\\
T_{n} &= 1 s\\
P_{2} &= 1.2\times 10^{-5}\ \\
p_{2z} &=  p_{2x} = p_{2y} = 10^{-4}\\
p_M &= 10^{-4}\\
s &= 5 \quad (\text{10 photons are used for measurement}).
\end{aligned}
\end{equation}
With these parameters, we have
\begin{equation}
\begin{aligned}
p_{I,M_z} &= 6.4\times 10^{-4}\\
p_{I,M_x} &= 5.4\times 10^{-4}\\
p_{I,M_y} &= 6.4\times 10^{-4}\\
p_{CZ} &= 5.4\times 10^{-3}.
\end{aligned}
\end{equation}
Error rates for measurement and initialisation are below our target error rate (0.1\%), but the CPHASE error rate above our target, but still below the actual threshold (0.73\%). The primary cause of this is the value of $P_2$. Recall that we assumed that a spin flip of the NV$^-$ electron always induces a phase flip on nucleus. In practice, the probability of an error depends quadratically on the exact fraction of time (relative to the 165 ns $\pi$ point of the hyperfine coupling) the system spends in the $\ket{+1}$ state, $\sin^2(t_{decay}\pi/165\;\textrm{ns})$.  The potential exists to engineer a much lower value of $P_2$ if the intrinsic branching ratios are similar to Fig.~\ref{fig:mixing}. For example, we could tune the cooperativity to decrease the scattering probability $P_S (0)$ for the $\ket{0} \leftrightarrow \ket{M_5}$ transition. Also, photons used in electron measurement could be sent at appropriate times to ensure that, in the case of absorption, a decay to the $\ket{+1}$ state occurs close to the $2\pi$ point of the hyperfine coupling. Reducing $P_2$ will be sufficient to reduce all error rates to below our target error rate.

\subsection{Expected performance}

Lastly, we estimate the performance of our architecture. The rate-limiting process is the connection of all electron--electron pairs in each step of the circuit to prepare the topological cluster state.  As discussed, we can operate the architecture in a synchronous or asynchronous manner, and the mode of operation will affect the performance. Simplest is the synchronous mode, where 99.9\% of all connections are established before moving to the next step (connections that are not established are introduce errors, which can be corrected). In this case, approximately 107 attempts per step are required for $p_c=6.25\%$.  This leads to a time per step of $(200\times 107+275) \approx 22$ $\mu$s.  In asynchronous mode, we take the average number of attempts for connections to be established ($1/p_c$). This implicitly assumes that different parts of the NV$^-$ array may by at different temporal stages of the computation, but classical control will be used to keep track of the entire topological cluster state, which is generated at a constant rate on average. In this case $\approx 16$ connection attempts are needed at $p_c = 6.25\%$ requiring a time of $3.5$ $\mu$s. Initialisation and measurement takes approximately 1 $\mu$s, so this is not the rate-limiting process. The quantum circuit illustrated in Fig.~\ref{fig:circuit} takes six steps to construct a layer of the topological cluster state.  1Hence, a temporal layer of the topological cluster state is prepared every $\approx 132$ $\mu$s in synchronous mode and every $\approx 21$ $\mu$s in asynchronous mode, with a unit cell prepared every $264$ $\mu$s and $42$ $\mu$s, respectively.

To estimate the size of topological cluster state and the speed of performing logical gate operations, we estimate the failure rate of a logical cell and the number of logical cells required for a logical gate \cite{RHG07}. The failure rate of a logical cell can be approximated as $p_L \approx C_1(C_2p/p_{th})^{(d+1)/2}$, where $d$ is the distance of the topological code, $p$ is the physical error rate, $p_{th}$ is the threshold error rate (estimated to be approximately 0.73\%), $C_1 \approx 0.13$, and $C_2 \approx 0.61$ \cite{fowler2011} We assume $p = 0.1\%$ is our average error rate for all gates, as the CPHASE gate has a slightly higher error and the measurement gates have slightly lower errors. For a large computation, we are likely to require $p_L \leq 10^{-18}$, implying $d \geq 32$. Then, a logical cell is a cube of unit cells measuring $5d/4 = 40$ cells in edge length. A logical qubit is defined as a cross section of the cluster, measuring $2\times 1$ logical cells, requiring $80\times 40$ unit cells. To perform a logical CNOT gate we require a cluster volume $2\times 2$ in cross section, requiring 9841 physical qubits and 2 logical cells in temporal depth. Hence, the time for a logical CNOT is $2\times 40\times 264 = 21.1$ ms for the synchronous mode and 3.4 ms for the asynchronous mode.


\begin{thebibliography}{99}
\bibitem{footnote1}
This is a fast and clean operation so long as a $\sigma_+$ polarized field is used. The polarized field stops population of the $| -1\rangle$ level.
\end{thebibliography}
\end{document}